\pdfoutput=1

\documentclass[11pt,twoside,a4paper,cmspaper,final,collab]{cms-tdr}
\def\svnVersion{238275}\def\cmsCernNo{2013-219}\def\cmsCernDate{\today}\def\cmsMessage{Published in the Journal of High Energy Physics as \href{http://dx.doi.org/10.1007/JHEP04(2014)103}{\doi{10.1007/JHEP04(2014)103.}}}

\begin{document}\cmsNoteHeader{HIN-13-003}

\hyphenation{had-ron-i-za-tion}
\hyphenation{cal-or-i-me-ter}
\hyphenation{de-vices}

\RCS$Revision: 235613 $
\RCS$HeadURL: svn+ssh://svn.cern.ch/reps/tdr2/papers/HIN-13-003/trunk/HIN-13-003.tex $
\RCS$Id: HIN-13-003.tex 235613 2014-04-07 12:50:08Z mironov $

\cmsNoteHeader{HIN-13-003} 
\providecommand{\qqbar}{\cPq\cPaq\xspace}
\newcommand{\QQbar}{\ensuremath{{\cmsSymbolFace{Q}\overline{\cmsSymbolFace{Q}}}}\xspace}
\renewcommand{\PgU}{\ensuremath{\Upsilon}\xspace}
\renewcommand{\PgUa}{\ensuremath{\Upsilon\text{(1S)}}\xspace}
\renewcommand{\PgUb}{\ensuremath{\Upsilon\text{(2S)}}\xspace}
\renewcommand{\PgUc}{\ensuremath{\Upsilon\text{(3S)}}\xspace}
\newcommand{\PgUbc}{\ensuremath{\Upsilon\text{(2S+3S)}}\xspace}
\newcommand{\PgUn}{\ensuremath{\Upsilon\text{(nS)}}\xspace}

\newcommand{\raa}{\ensuremath{R_\mathrm{AA}}\xspace}

\newcommand{\pp}{{\ensuremath{\Pp\Pp}}\xspace}
\newcommand{\ppbar}{\ensuremath{\Pp\Pap}\xspace}
\newcommand{\PbPb}{\ensuremath{\text{PbPb}}\xspace}
\newcommand{\pPb}{\ensuremath{\Pp\text{Pb}}\xspace}

\newcommand{\sqrts}{\ensuremath{\sqrt{s}}\xspace}
\newcommand{\sqrtsnn}{\ensuremath{\sqrt{s_{_{\text{NN}}}}}\xspace}

\providecommand{\mbinv} {\mbox{\ensuremath{\,\text{mb}^\text{$-$1}}}\xspace}

\newcommand{\doubleRatioUps}{\ensuremath{[\PgUb,\PgUc/\PgUa]_{\text{\pPb}}/[\PgUb,\PgUc/\PgUa]_{\text{\pp}}}\xspace}
\newcommand{\doubleRatioUpsBA}{\ensuremath{[\Upsilon\text{(2S)}/\Upsilon\text{(1S)}]_{\text{\pPb}}/[\Upsilon\text{(2S)}/\Upsilon\text{(1S)}]_{\text{\pp}}}\xspace}
\newcommand{\doubleRatioUpsCA}{\ensuremath{[\Upsilon\text{(3S)}/\Upsilon\text{(1S)}]_{\text{\pPb}}/[\Upsilon\text{(3S)}/\Upsilon\text{(1S)}]_{\text{\pp}}}\xspace}

\newcommand{\singleRatioUpsXA}{\ensuremath{\Upsilon\text{(nS)}/\Upsilon\text{(1S)}}\xspace}
\newcommand{\singleRatioUpsBA}{\ensuremath{\PgUb/\PgUa}\xspace}
\newcommand{\singleRatioUpsCA}{\ensuremath{\PgUc/\PgUa}\xspace}

\newcommand{\HFetaRelative}{\ensuremath{E_{\text{T}}^{\abs{\eta}>4}/{\langle E_{\text{T}}^{\abs{\eta}>4}\rangle}_{\text{total}}}\xspace}
\newcommand{\HFeta}{\ensuremath{E_{\mathrm{T}}^{\abs{\eta}>4}}\xspace}

\newcommand{\NtrksRelative}{\ensuremath{N_{\text{tracks}}^{\abs{\eta}<2.4}/{\langle N_{\text{tracks}}^{\abs{\eta}<2.4}\rangle}_{\text{total}}}\xspace}
\newcommand{\Ntrks}{\ensuremath{N_{\text{tracks}}^{\abs{\eta}<2.4}}\xspace}

\providecommand{\CASCADE} {{\textsc{cascade}}\xspace}
\providecommand{\HYDJET} {{\textsc{hydjet}}\xspace}

\title{\texorpdfstring{Event activity dependence of $\Upsilon$(nS) production in $\sqrt{s_{NN}} = 5.02\TeV$ \pPb and $\sqrt{s} = 2.76\TeV$ pp collisions}{Event activity dependence of Y(nS) production in sqrt(s[NN])=5.02 TeV pPb and sqrt(s)=2.76 TeV pp collisions}}

\date{\today}

\abstract{The production of $\Upsilon$(1S), $\Upsilon$(2S), and $\Upsilon$(3S) is investigated in pPb and pp collisions at centre-of-mass energies per nucleon pair of 5.02\TeV and 2.76\TeV, respectively. The datasets correspond to integrated luminosities of about 31\nbinv (pPb) and 5.4\pbinv (pp), collected in 2013 by the CMS experiment at the LHC. Upsilons that decay into muons are reconstructed within the rapidity interval $\abs{y_{\mathrm{CM}}}<1.93$ in the nucleon-nucleon centre-of-mass frame. Their production is studied as a function of two measures of event activity, namely the charged-particle multiplicity measured in the pseudorapidity interval $\abs{\eta}<2.4$, and the sum of transverse energy deposited at forward pseudorapidity, $4.0<\abs{\eta}<5.2$. The $\Upsilon$ cross sections normalized by their event activity integrated values, $\Upsilon\text{(nS)}/\langle\Upsilon\text{(nS)}\rangle$, are found to rise with both measures of the event activity in pp and pPb. In both collision systems, the ratios of the excited to the ground state cross sections, $\Upsilon\text{(nS)}/\Upsilon\text{(1S)}$, are found to decrease with the charged-particle multiplicity, while as a function of the transverse energy the variation is less pronounced. The event activity integrated double ratios, $[\Upsilon\text{(nS)}/\Upsilon\text{(1S)}]_{\pPb} / [\Upsilon\text{(nS)}/\Upsilon\text{(1S)}]_{\pp}$, are also measured and found to be $0.83 \pm 0.05\stat\pm 0.05\syst$ and $0.71 \pm 0.08\stat\pm 0.09\syst$ for $\Upsilon\text{(2S)}$ and $\Upsilon\text{(3S)}$, respectively.}

\hypersetup{%
pdfauthor={CMS Collaboration},%
pdftitle={Event activity dependence of Y(nS) production in sqrt(s[NN])=5.02 TeV pPb and sqrt(s)=2.76 TeV pp collisions},%
pdfsubject={CMS},%
pdfkeywords={CMS, physics, heavy-ions, pp, pPb, dimuons, quarkonia}
}

\maketitle 

\section{Introduction}
The suppression of the \PgUa, \PgUb, and \PgUc (collectively referred to as \PgUn in what follows) yields produced in heavy-ion collisions relative to proton-proton (pp) collisions was first measured by the  Compact Muon Solenoid (CMS) experiment, at the Large Hadron Collider (LHC), in \PbPb collisions at a centre-of-mass energy per nucleon pair of $\sqrtsnn = 2.76$\TeV~\cite{Chatrchyan:2011pe, Chatrchyan:2012lxa}. The tightest bound state, \PgUa, was observed to be less suppressed than the more loosely bound excited states, \PgUb and \PgUc. Such ordering is theoretically predicted to occur in the presence of a deconfined medium in which the colour fields modify the spectral properties of the \bbbar quark pair, and prevent the formation of a bound state~\cite{Matsui:1986dk, Digal:2001ue,Mocsy:2007jz,Laine:2006ns}. However, other phenomena, discussed below, can affect the bottomonium yields at stages that precede or follow the formation of the \bbbar pair and of the bound state, independently of the presence of a deconfined partonic medium. Some of these phenomena could lead to a suppression sequence that depends on the binding energy. In this context, measurements in reference systems are essential: proton-lead (\pPb) collisions can probe nuclear effects, while pp collisions are essential for understanding the elementary bottomonium production mechanisms.

In heavy-ion collisions (AA), effects that precede the formation of the \bbbar pair (called here initial-state effects), such as the modification of the nuclear parton distribution functions (nPDFs) in the incoming nuclei~\cite{Vogt:2010aa}, parton energy loss, and the Cronin effect~\cite{Sharma:2012dy,Arleo:2012rs}, are expected to affect the members of the \PgU family in the same way, given their small mass difference and identical quantum numbers $J^{\mathrm{CP}}=1^{--}$. Consequently, any difference among the states is likely due to phenomena occurring after the \bbbar production, during or after the \PgU formation. Examples of final-state effects that might play a role include interactions with spectator nucleons that break up the state (nuclear absorption)~\cite{Gerschel:1988wn,Lourenco:2008sk}, and collisions with comoving hadrons~\cite{Gavin:1990gm,Capella:2007jv} or surrounding partons~\cite{Brodsky:1988xz,Laine:2006ns,Strickland:2011aa,Emerick:2011xu,Sharma:2012dy} that can dissociate the bound states or change their kinematics. Any of these final-state processes can affect the \PgUn yields differently, depending on the binding energy and size of each state, and be at play in AA and/or pA collisions, possibly with different strengths and weights, depending on the properties of the environment created in each case. A measurement of the \PgUa and \PgUbc production cross sections in pA collisions at $\sqrtsnn \approx 39$\GeV using several targets, relative to proton-deuterium collisions~\cite{Alde:1991sw}, showed no difference, within uncertainties, between the ground state and the combined excited states, although a suppression was observed for both.

Understanding the production of bottomonia in elementary pp collisions is equally important for interpreting any additional effects in collisions involving heavy ions. At present, there are different proposed mechanisms to describe the evolution of a heavy-quark pair into a bound quarkonium state (a review can be found in e.g. Ref.~\cite{Brambilla:2010cs}), but little is known of the underlying event associated with each state. For instance, the fragmentation of the soft gluons involved in some mechanisms~\cite{Beneke:1996tk,Vogt:2001ky}, or the feed-down processes~\cite{Digal:2001ue} (decays of the higher-mass states to one of lower mass) could generate different numbers of particles associated with each of the quarkonium states. Therefore, the average contribution from each state to the global event characteristics (multiplicity, transverse energy, etc) can be different. In addition, the recent observation in pp collisions at $\sqrts=7$\TeV~\cite{Abelev:2012rz} that the \JPsi yield increases with associated track multiplicity suggests that other phenomena need to be considered for a full understanding of the quarkonium production mechanism in elementary collisions.

This paper reports measurements of three observables characterizing the \PgU mesons produced in \pp and \pPb collisions within the interval $\abs{y_{\mathrm{CM}}}<1.93$, where $y_{\mathrm{CM}}$ is the meson rapidity in the centre-of-mass of the nucleon-nucleon collision. First, double ratios of the yields of the excited states, \PgUb and \PgUc, to that of the ground state, \PgUa, are reported in \pPb with respect to pp collisions, \doubleRatioUpsBA, and similarly for the \PgUc. Then, single yield ratios of the excited states to the ground state, \singleRatioUpsXA, are corrected for detector acceptance and reconstruction inefficiencies, and studied as a function of two event activity variables, measured in different rapidity ranges: a) the sum of the transverse energy deposited at a large rapidity gap with respect to the \PgU, in the forward region ($4.0<\abs{\eta}<5.2$), and b) the number of charged particles reconstructed in the central region ($\abs{\eta}<2.4$) that includes the rapidity range in which the $\Upsilon$ is measured. Lastly, \PgUn cross sections are studied as a function of the same event activity variables, with both cross sections and event activities divided by their values in all measured events. These values (denoted "activity-integrated values") are found by including all events with no selection on transverse energy or particle multiplicity.

\section{Experimental setup and event selection}
The results presented in this paper use pp data corresponding to an integrated luminosity of 5.4\pbinv, and \pPb collision data corresponding to an integrated luminosity of $31 \nbinv$. The pp data were collected at a centre-of-mass energy $\sqrts =2.76$\TeV. In \pPb collisions the beam energies were 4\TeV for protons, and 1.58\TeV per nucleon for lead nuclei, resulting in a centre-of-mass energy per nucleon pair of $\sqrtsnn = 5.02$\TeV. The direction of the higher-energy proton beam was initially set up to be clockwise, and was reversed after an integrated luminosity of 18\nbinv of data was recorded. As a result of the energy difference of the colliding beams, the nucleon-nucleon centre-of-mass in the \pPb collisions is not at rest with respect to the laboratory frame. Massless particles emitted at $\abs{\eta_{\mathrm{CM}}}$ = 0 in the nucleon-nucleon centre-of-mass frame are detected at $\eta = -0.465$ (clockwise proton beam) or $+0.465$ (counterclockwise proton beam) in the laboratory frame.

A detailed description of the CMS detector can be found in Ref.~\cite{CMS:2008zzk}. Its main feature is a superconducting solenoid of 6\unit{m} internal diameter, providing a magnetic field of 3.8\unit{T}. Within the field volume are the silicon  pixel and strip tracker, the crystal electromagnetic calorimeter, and the brass/scintillator hadron calorimeter. The silicon pixel and strip tracker measures charged-particle trajectories in the  range $\abs{\eta} < 2.5$. It consists of 66\unit{M} pixel and 10\unit{M} strip channels.
Muons are detected in the range $\abs{\eta} < 2.4$, with detection planes based on three technologies: drift tubes, cathode strip chambers, and resistive-plate chambers. Because of the strong magnetic field and the fine granularity of the tracker, the muon \pt measurement based on information from the tracker alone has a resolution between 1\% and 2\% for a typical muon in this analysis. The CMS apparatus also has extensive forward calorimetry, including two steel/quartz-fibre Cherenkov hadron forward (HF) calorimeters, which cover the range $2.9 < \abs{\eta} < 5.2$. These forward calorimeters are used for online event selection and provide a measure of the forward event activity.

Similar selection criteria as the ones developed in Ref.~\cite{Chatrchyan:2013nka} are applied to the \pPb sample to remove electromagnetic, beam-gas, and multiple collisions (pileup). The longitudinal and transverse distance between the leading vertex (the vertex with the highest number of associated tracks) and the second vertex in an event are used as criteria for identifying and removing pileup events. These criteria are tightened when applied to the pp sample, which has a higher number of simultaneous collisions per beam crossing; at maximum, at the beginning of an LHC fill, 23\% of the pp events had more than one collision, compared to 3\% in \pPb. After the selection, the remaining integrated luminosity in the \pp sample is equivalent to 4.1\pbinv, with a residual pileup lower than 3\%.  Since pileup only biases the event activity variables, this selection is applied to the event activity dependent part of the analysis, but not for the \pp integrated results.

Monte Carlo (MC) events are used to evaluate efficiencies and acceptances. Signal \PgUn events are generated, for 2.76\TeV and 5.02\TeV (boosted to have the correct rapidity distribution in the detector frame), using \PYTHIA 6.424~\cite{Sjostrand:2006za}. In all samples, the \PgUn decay is simulated using \EVTGEN~\cite{Lange:2001uf}, assuming unpolarized production~\cite{Chatrchyan:2012woa}. No systematic uncertainties are assigned for this assumption, any possible modification due to polarization being considered as part of the physics that is studied~\cite{Faccioli:2012kp}. The final-state bremsstrahlung is implemented using \PHOTOS~\cite{Barberio:1993qi}. The CMS detector response is simulated with \GEANTfour~\cite{Agostinelli:2002hh}.

\section{Signal extraction}
\begin{figure}[t]
  \begin{center}
    \includegraphics[width=0.45\textwidth]{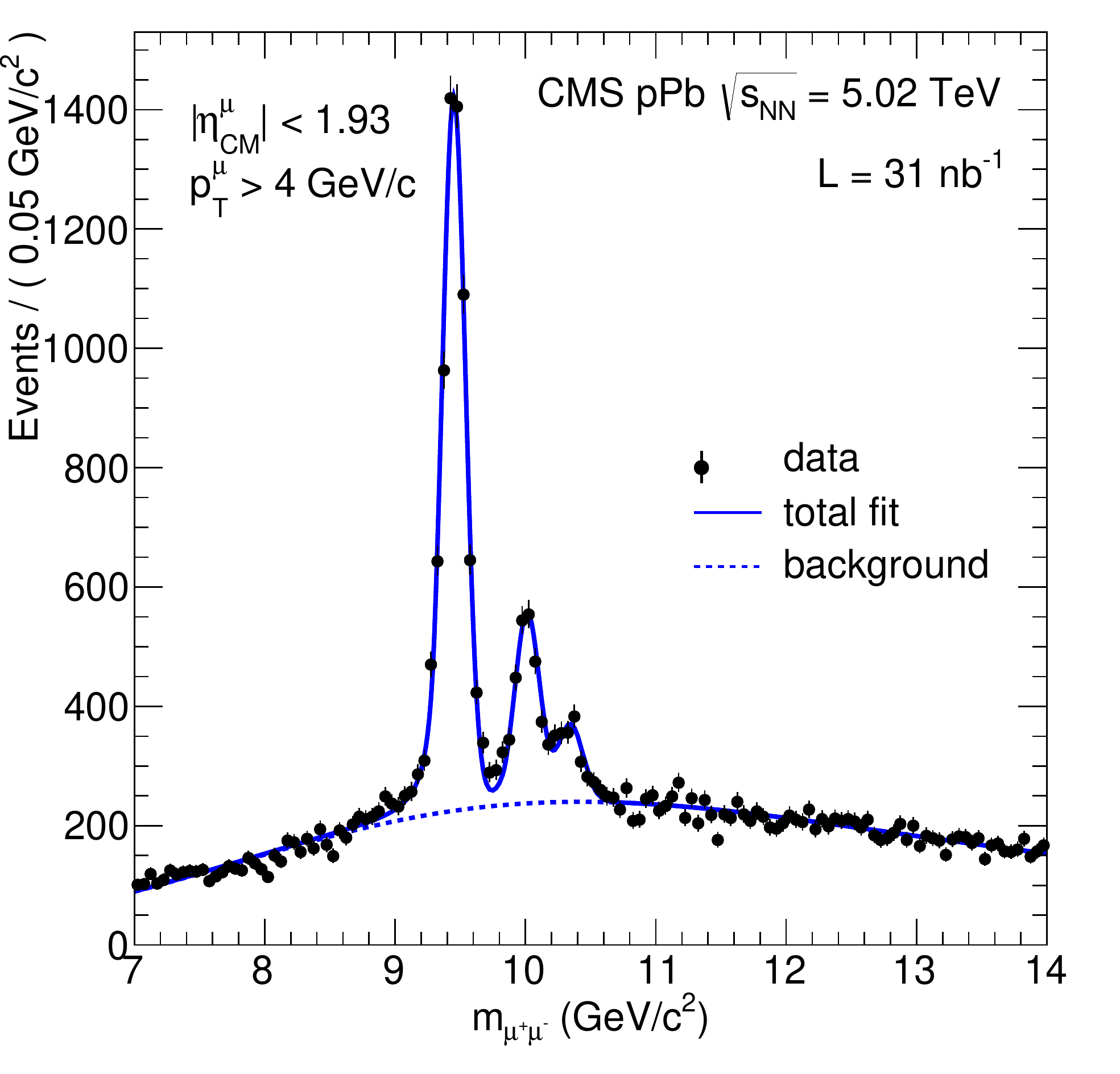}\hspace{1em}
    \includegraphics[width=0.45\textwidth]{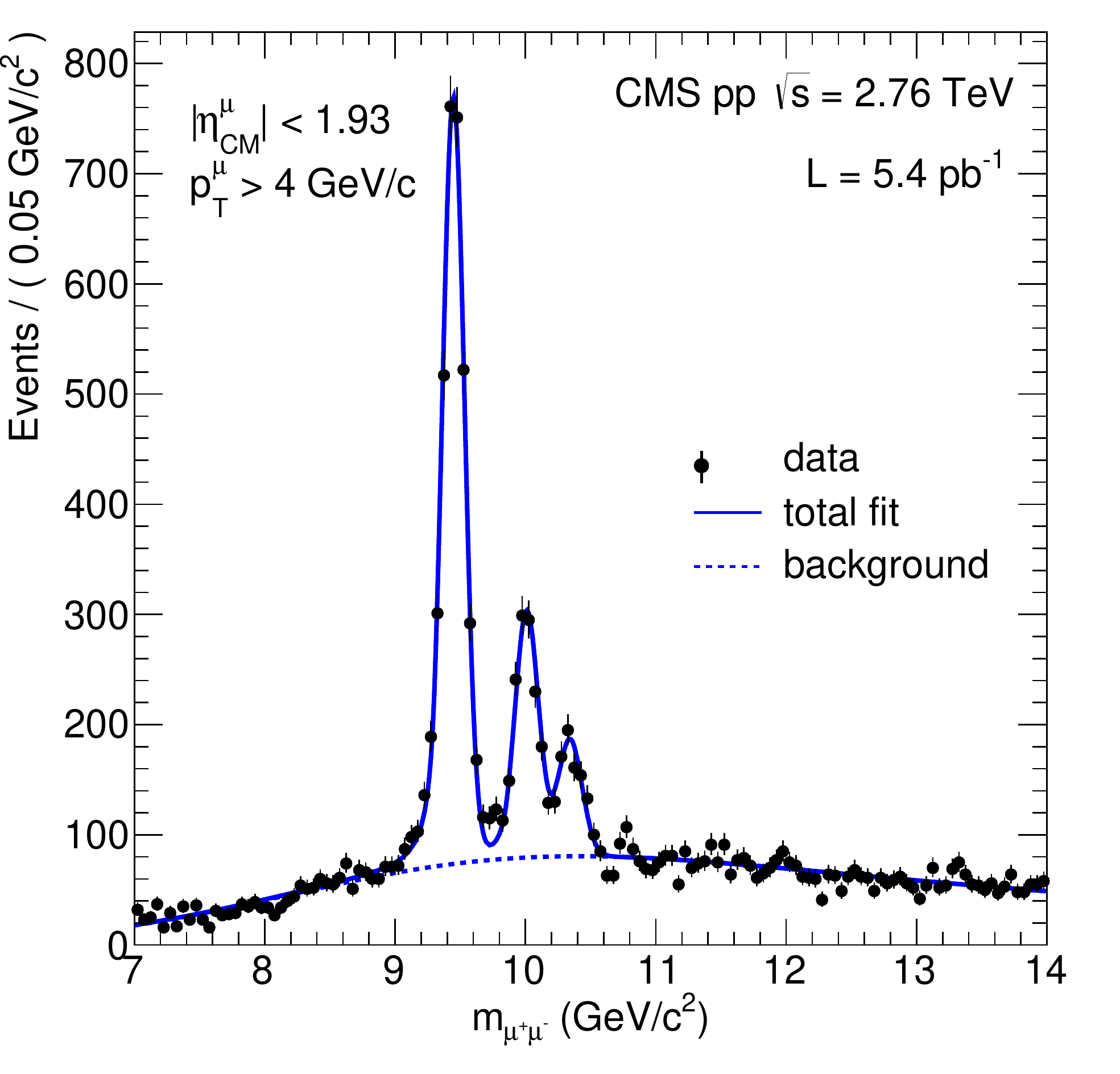}\\
    \caption{Invariant mass spectrum in \pPb (left) and pp collisions
      (right) of \Pgmp\Pgmm\ pairs with single muons with  $\pt^{\mu}>4$\GeVc and $\abs{\eta^\mu_{\mathrm{CM}}}<1.93$. The data (black circles) are overlaid with the fit (solid blue line). The background component of the fit is represented by the dashed blue line.}
    \label{fig:massFits}
  \end{center}
\end{figure}

The $\PgU$ states are identified through their dimuon decay. The events were selected online with a hardware-based trigger requiring two muon candidates in the muon detectors with no explicit momentum or rapidity thresholds. Offline, only reconstructed muons with pseudorapidity $\abs{\eta^\mu_{\mathrm{CM}}}<1.93$ and transverse momentum $\pt^\mu > 4$\GeVc, passing the quality requirements described in Ref.~\cite{Khachatryan:2010zg}, are selected. The $\pt^\mu$ selection is identical to the one used in the PbPb analyses~\cite{Chatrchyan:2011pe, Chatrchyan:2012lxa}, but the individual muon $\abs{\eta^\mu_{\mathrm{CM}}}$ is restricted to be smaller than 1.93, in order to keep a symmetric range in the \pPb centre-of-mass frame. The same selections are used when analyzing the \pPb and \pp data. The \pt range of the selected dimuon candidates extends down to zero. The dimuon rapidity is limited to $\abs{y_{\mathrm{CM}}}<1.93$. The resulting opposite-charge dimuon invariant-mass distributions are shown in Fig.~\ref{fig:massFits} for the \pPb (left) and \pp (right) datasets, in the 7--14\GeVcc range.

The \singleRatioUpsXA yield ratios are extracted from an unbinned maximum likelihood fit to the invariant dimuon mass spectra, following the method described in Ref.~\cite{Chatrchyan:2012lxa}. The reconstructed mass lineshape of each $\PgUn$ state is modeled by a Crystal Ball (CB) function~\cite{Skwarnicki:1986xj}, \ie a Gaussian function with the low-side tail replaced by a power law function describing final-state radiation. The mass resolution, described by the width of the Gaussian component of the CB, is constrained to scale with the ratios of the resonance masses. The resolution of the \PgUa mass is a free parameter in the activity-integrated fits, and fixed to the value obtained in the integrated fits when fitting in bins of event activity. Reasonable variations  with multiplicity are considered in the systematic uncertainties. The CB tail parameters are fixed to values obtained from MC simulations. The \PgUn mass ratios are fixed to their world average values~\cite{Beringer:1900zz}, with the \PgUa mass left free and found to be consistent with its world average value. The background shape is modeled by an exponential function multiplied by an error function and all its parameters are left free in the fit, as in Ref.~\cite{Chatrchyan:2012lxa}.

The systematic uncertainties from the signal extraction are evaluated by allowing different line-shape variations. The signal shape is varied by fixing all CB parameters to their MC expectations, fixing only one CB parameter to the expectation, and leaving all CB parameters floating free. The background model is varied by using different shapes, and by constraining its parameters from a fit to the same-sign dimuon spectrum. The maximum observed variations are taken as a conservative estimate of the corresponding systematic uncertainties.

The pp reference data are taken at a different nucleon-nucleon centre-of-mass energy than the \pPb data. In order to assess the \sqrts dependence of the single ratio in pp collisions, the single ratios measured at $\sqrts = 7$\TeV~\cite{Chatrchyan:2013yna} and $\sqrts = 1.8$\TeV~\cite{Abe:1995an,Acosta:2001gv}, tabulated in Table~\ref{tab:sqrts}, are compared to the $\sqrts = 2.76$\TeV ratios of the present analysis. No significant difference is found within the systematic and statistical uncertainties in all samples. The 2.76\TeV pp sample is used to compute the double ratios since it was recorded with the same trigger requirements and reconstructed with the same algorithms as the \pPb data, and hence the related efficiencies cancel in the double ratio, down to a level which is negligible ($<$0.1\%) with respect to other systematic and statistical uncertainties. It is further checked, for each sample, that the trigger, reconstruction, and selection efficiencies agree well, to better than 2\%, between data and simulations (following the same procedure as in Ref.~\cite{Chatrchyan:2012np}).

\section{Event activity integrated results}
\subsection{Double ratios: \texorpdfstring{$[\Upsilon(\text{nS})/\Upsilon(\text{1S})]_{\text{pPb}} /  [\Upsilon(\text{nS})/\Upsilon(\text{1S})]_{\text{pp}}$}{dblRatio}}
Using the raw yield ratios found by fitting separately the \pPb and pp event activity integrated data samples, the double ratios are
\begin{align*}
\label{eq:double}
  \frac{\PgUb/\PgUa\vert_{\pPb}}{\PgUb/\PgUa\vert_{\pp}} & =   0.83 \pm 0.05\stat\pm 0.05\syst\\
  \frac{\PgUc/\PgUa\vert_{\pPb}}{\PgUc/\PgUa\vert_{\pp}} & =   0.71 \pm 0.08\stat \pm 0.09\syst.
\end{align*}

The systematic uncertainties include uncertainties from the signal extraction procedure described above (6\% and 13\% for the \PgUb and \PgUc, respectively), and from a potentially imperfect cancelation of the acceptances for individual states between the two centre-of-mass energies (2\% and 1\%, respectively, estimated from MC).

The above double ratios, in which the initial-state effects are likely to cancel, suggest the presence of final-state effects in the \pPb collisions compared to pp collisions, that affect more strongly the excited states (\PgUb and \PgUc) compared to the ground state (\PgUa).

\begin{figure}[t]
  \begin{center}
     \includegraphics[width=0.45\textwidth]{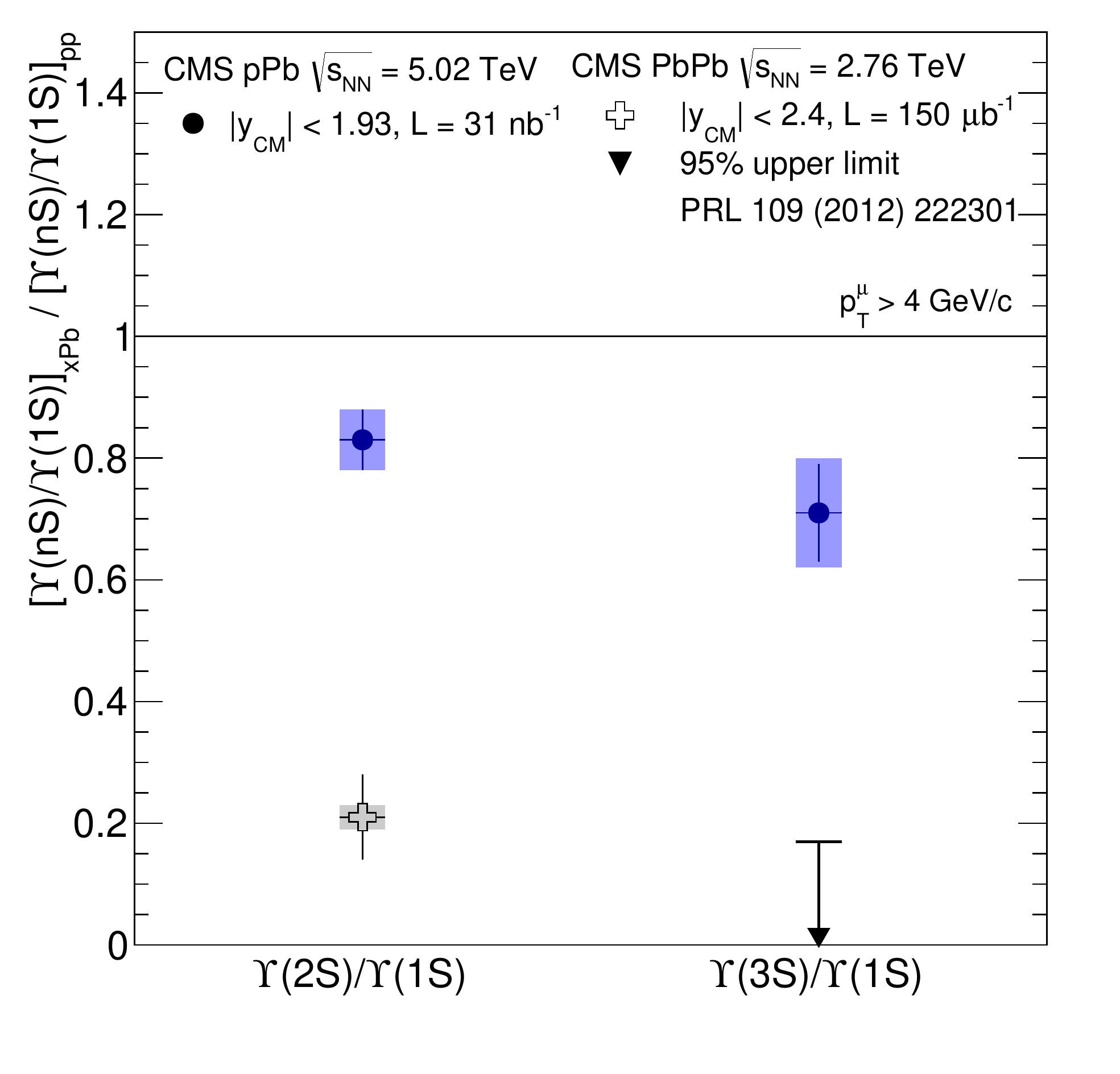}
      \includegraphics[width=0.45\textwidth]{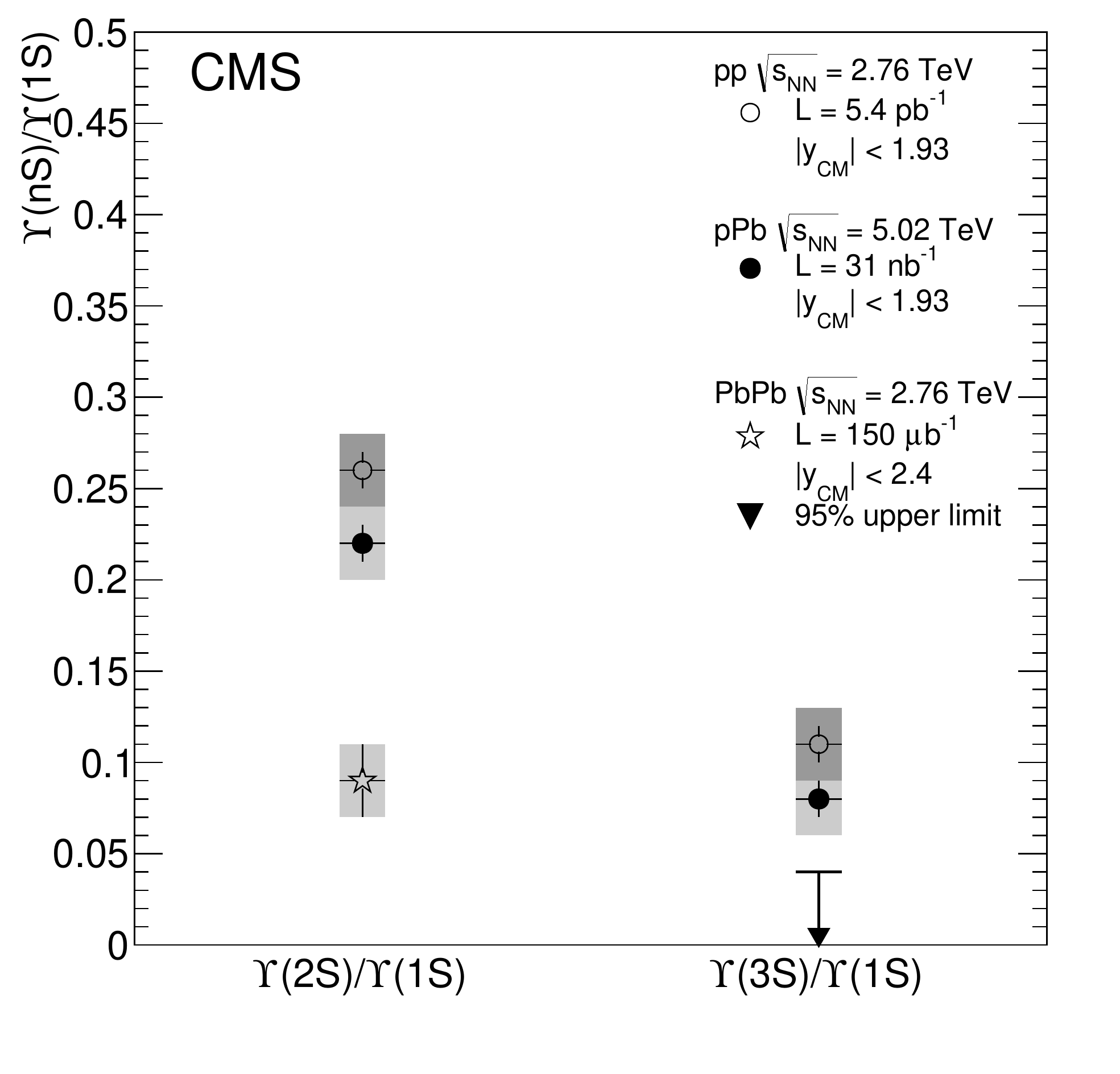}
    \caption{Left: event activity integrated double ratios of the excited states, \PgUb and \PgUc, to the ground state,\PgUa, in \pPb collisions at $\sqrtsnn=5.02$\TeV with respect to pp collisions at $\sqrts=2.76$\TeV (circles), compared to the corresponding ratios for \PbPb (cross) collisions at $\sqrtsnn=2.76$\TeV from Ref.~\cite{Chatrchyan:2012lxa}, which used a different dataset for the pp normalization. Right: event activity integrated single cross section ratios of the excited \PgUb and \PgUc to the ground \PgUa state, as measured in pp (open circles), \pPb (full circles), and \PbPb (open star) collisions at 2.76, 5.02, and 2.76\TeV, respectively. In both figures, the error bars indicate the statistical uncertainties, and the boxes represent the systematic uncertainties. The single ratios are available in tabulated form in Table~\ref{tab:fig2left}.}
    \label{fig:dblRatio}
  \end{center}
\end{figure}

In Fig.~\ref{fig:dblRatio} (left), the \pPb double ratios are compared with the measurement in \PbPb at $\sqrtsnn = 2.76$\TeV~\cite{Chatrchyan:2012lxa}. The \pPb ratios are larger than the corresponding \PbPb ones. This observation may help in understanding the final-state mechanisms of suppression of excited $\Upsilon$ states in the absence of a deconfined medium, and their extrapolation to the \PbPb system. It is noted here that the PbPb double ratios reported in Ref.~\cite{Chatrchyan:2012lxa} were normalized to the smaller pp dataset collected by CMS in 2011. Once all the corrections are applied, the ratio of the 2011 to the 2013 pp single cross section ratios is $1.6 \pm 0.4\stat$, making them consistent within 1.5 standard deviations. Normalizing by the 2013 reference data would bring the PbPb double ratio up by the same factor 1.6 and reduce the statistical uncertainties, at the price of enhancing the systematic uncertainties since the trigger and reconstruction algorithm are different. Also, though single ratios in pp collisions do not depend significantly on $\sqrts$~\cite{Chatrchyan:2013yna,Abe:1995an} and on rapidity~\cite{Chatrchyan:2013yna}, one should take into account when comparing or extrapolating the results in Fig.~\ref{fig:dblRatio} that the \pPb and \PbPb single ratios differ in these aspects.

\subsection{Single cross section ratios: \texorpdfstring{$\Upsilon(\text{nS})/\Upsilon(\text{1S})$}{excitesUpsilons1S}}
The single ratios used as numerator and denominator in the \pPb double ratios in Fig.~\ref{fig:dblRatio} (left) are further corrected for detector acceptance (to a single muon transverse momentum coverage of $\pt^{\mu}>0$\GeVc and Upsilon $\abs{y_{\mathrm{CM}}}<1.93$), reconstruction and trigger inefficiencies, and are given in Fig.~\ref{fig:dblRatio} (right). The global uncertainties (not related to the signal extraction) are added in quadrature to the systematic uncertainties, and are estimated by following the same methods as in the previous analyses~\cite{Chatrchyan:2012np,Chatrchyan:2012lxa}: by considering the effect of variations in the simulated kinematic distributions on the acceptance (7--8\%) and efficiency (1--2\%) corrections, and from differences in the efficiency estimations from data and MC simulation ($<1\%$). The \PbPb values are derived from Ref.~\cite{Chatrchyan:2012lxa} but, unlike the ones quoted in Eq.~(1) in that reference, they are corrected for acceptance and efficiency, following the same procedures as used for the 2013 samples. 

Similar to the double ratios, the single ratios signal the presence of different (or stronger) final state effects acting on the excited states compared to the ground state from \pp to \pPb to \PbPb collisions. For both types of ratios, a quantitative  extrapolation of these effects in \pPb to the corresponding \PbPb requires theoretical modeling, which is beyond the scope of this paper.

\section{Event activity binned results}
\subsection{Excited-to-ground state cross section ratios: \texorpdfstring{$\Upsilon(\text{nS})/\Upsilon(\text{1S})$}{binExciteUps1S}}
The pp and \pPb data are further analyzed separately as a function of event activity variables measured in two different rapidity regions. Specifically, the single ratios, \singleRatioUpsBA and \singleRatioUpsCA, are measured in bins of: (1) \HFeta, the raw transverse energy deposited in the most forward part of the HF calorimeters at $4.0<\abs{\eta}<5.2$, and (2) \Ntrks, the number of charged particles, not including the two muons, with $\pt>400$\MeVc reconstructed in the tracker at $\abs{\eta}<2.4$ and originating from the same vertex as the $\Upsilon$.

The binning is chosen using a minimum bias event sample, triggered by requiring at least one track with $\pt>400$\MeVc to be found in the pixel tracker for a bunch crossing. The bin upper boundaries, presented in Table~\ref{Tab:hfNtrkBin}, are chosen for each variable so that they are half or round multiples of the uncorrected mean value in the minimum bias events, $ {\langle N_{\text{tracks, raw}}^{\abs{\eta}<2.4}\rangle}= 10$ and 41, ${\langle E_{\text{T, raw}}^{\abs{\eta}>4} \rangle}= 3.5$ and 14.7\GeV for pp and \pPb, respectively. Table~\ref{Tab:hfNtrkBin} also lists, for each bin, the mean values of both variables, as computed from the dimuon sample used in the analysis, and the fraction of minimum bias events in the bin. For \Ntrks, the mean is extracted after weighting each reconstructed track in one bin by a correction factor that accounts for the detector acceptance, the efficiency of the track reconstruction algorithm, and the fraction of misreconstructed tracks as described in Ref.~\cite{Chatrchyan:2013nka}. Based on studies in Refs.~\cite{TRK-10-002, Chatrchyan:2012ta}, the uncertainty in the total single-track correction is estimated to be 3.9\% for the 2013 \pp and \pPb data, and 10\% for the \PbPb data.

\begin{figure}[t]
  \begin{center}
    \includegraphics[width=0.45\textwidth]{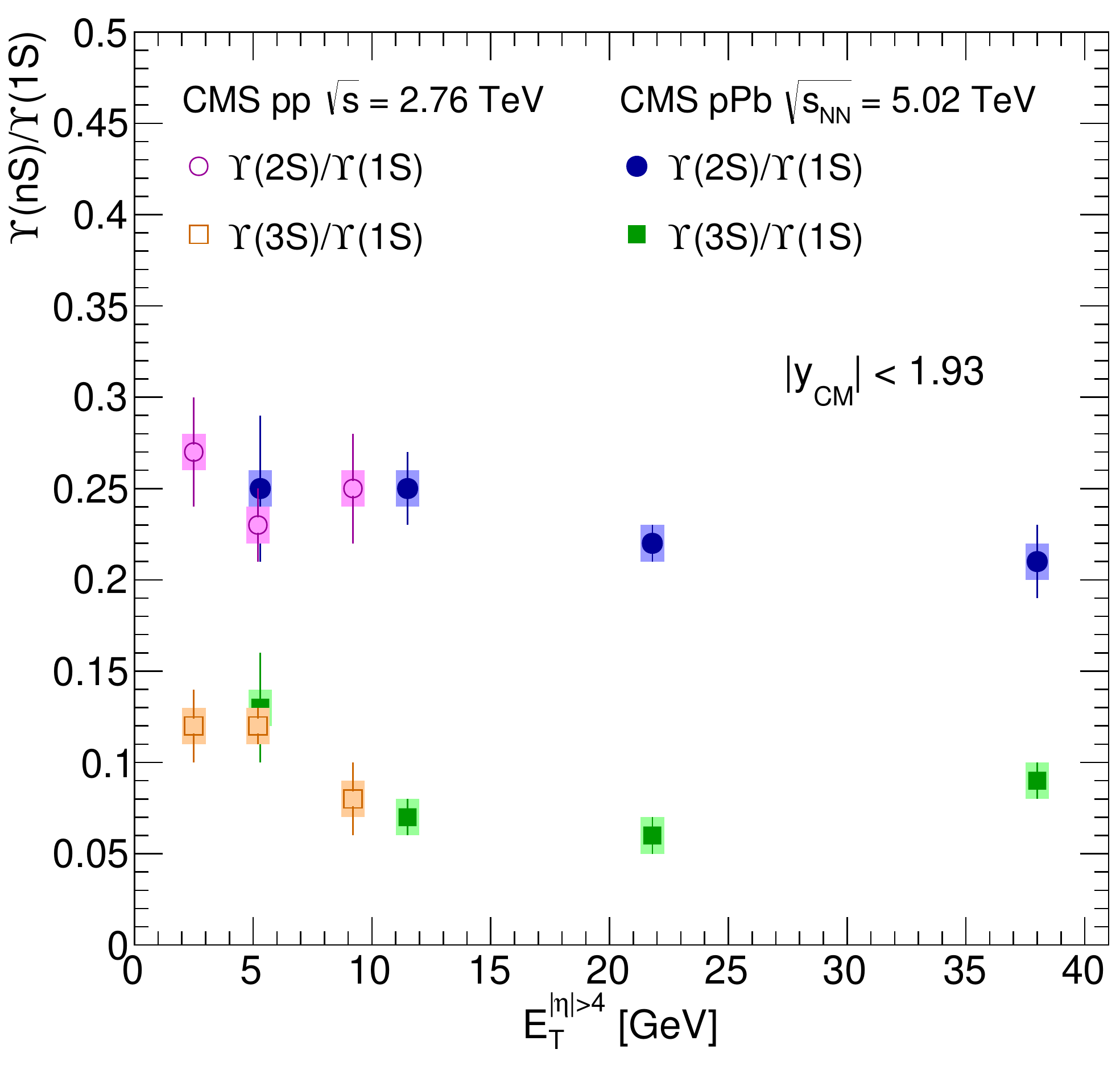}
   	     \includegraphics[width=0.45\textwidth]{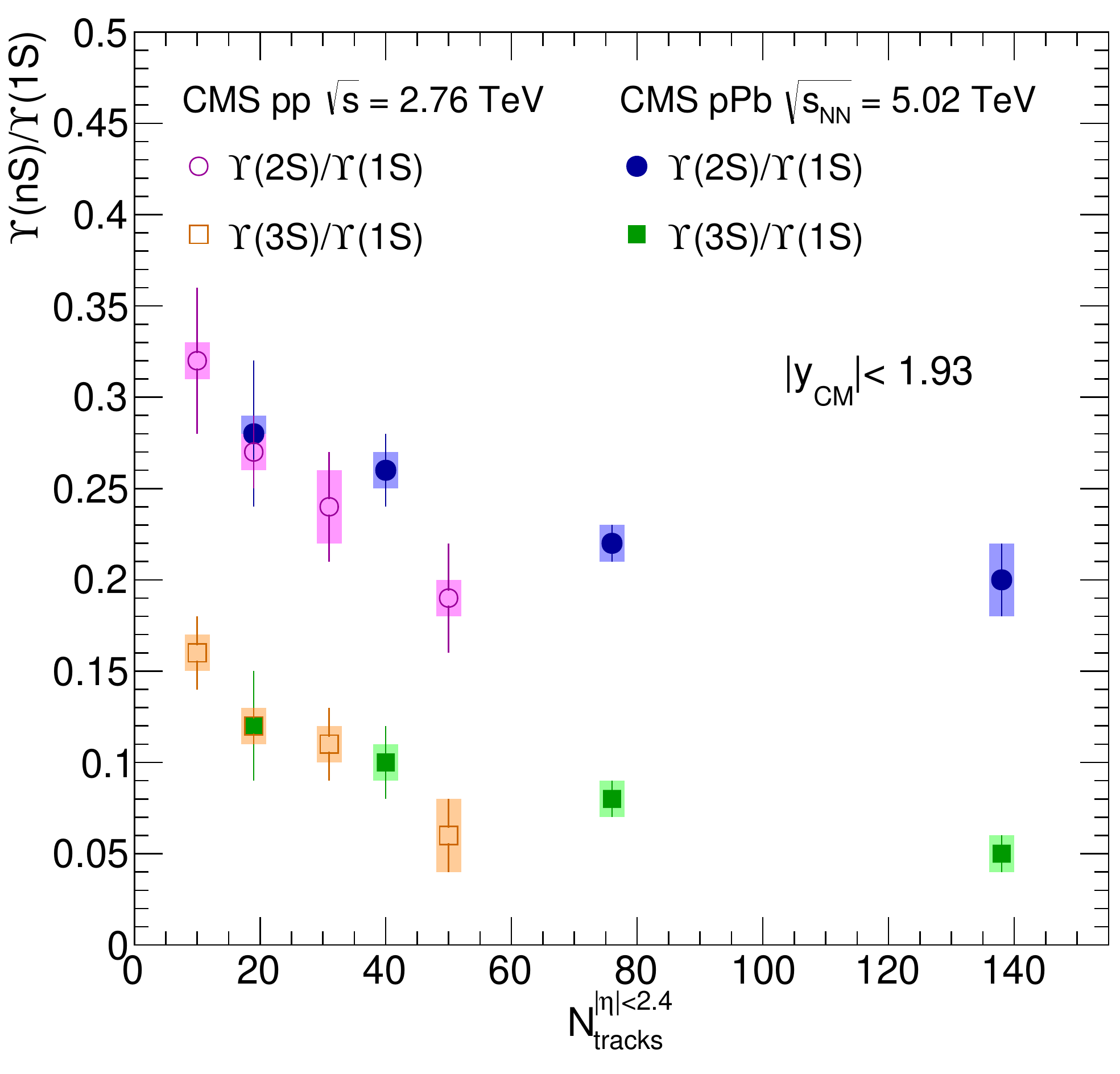}	
    \caption{Single cross section ratios \singleRatioUpsBA and \singleRatioUpsCA for $\abs{y_{\mathrm{CM}}}<1.93$ versus  transverse energy measured in $4.0<\abs{\eta}<5.2$ (left) and number of charged tracks measured in $\abs{\eta}<2.4$ (right), for pp collisions at $\sqrts=2.76$\TeV (open symbols) and \pPb collisions at $\sqrtsnn=5.02$\TeV (closed symbols). In both figures, the error bars indicate the statistical uncertainties, and the boxes represent the point-to-point systematic uncertainties. The global uncertainties on the \pp results are 7\% and 8\% for \singleRatioUpsBA and \singleRatioUpsCA, respectively, while in the \pPb results they amount to 8\% and 9\%, respectively. The results are available in tabulated form in Table~\ref{tab:fig3}, with binning information provided in Table~\ref{Tab:hfNtrkBin}.}
    \label{fig:ratio}
  \end{center}
\end{figure}

\begin{figure}[t]
  \begin{center}
    \includegraphics[width=0.45\textwidth]{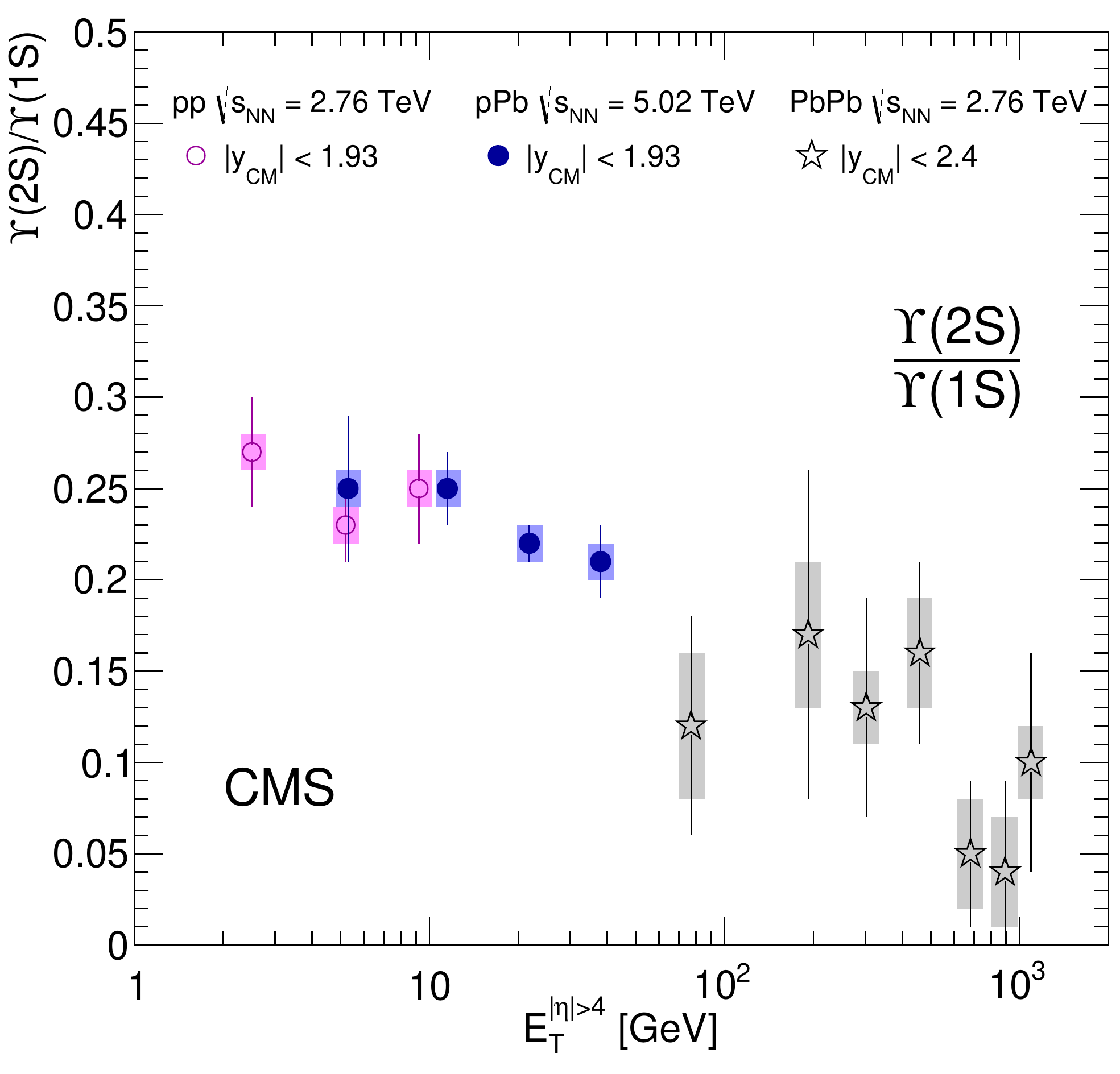}
    \includegraphics[width=0.45\textwidth]{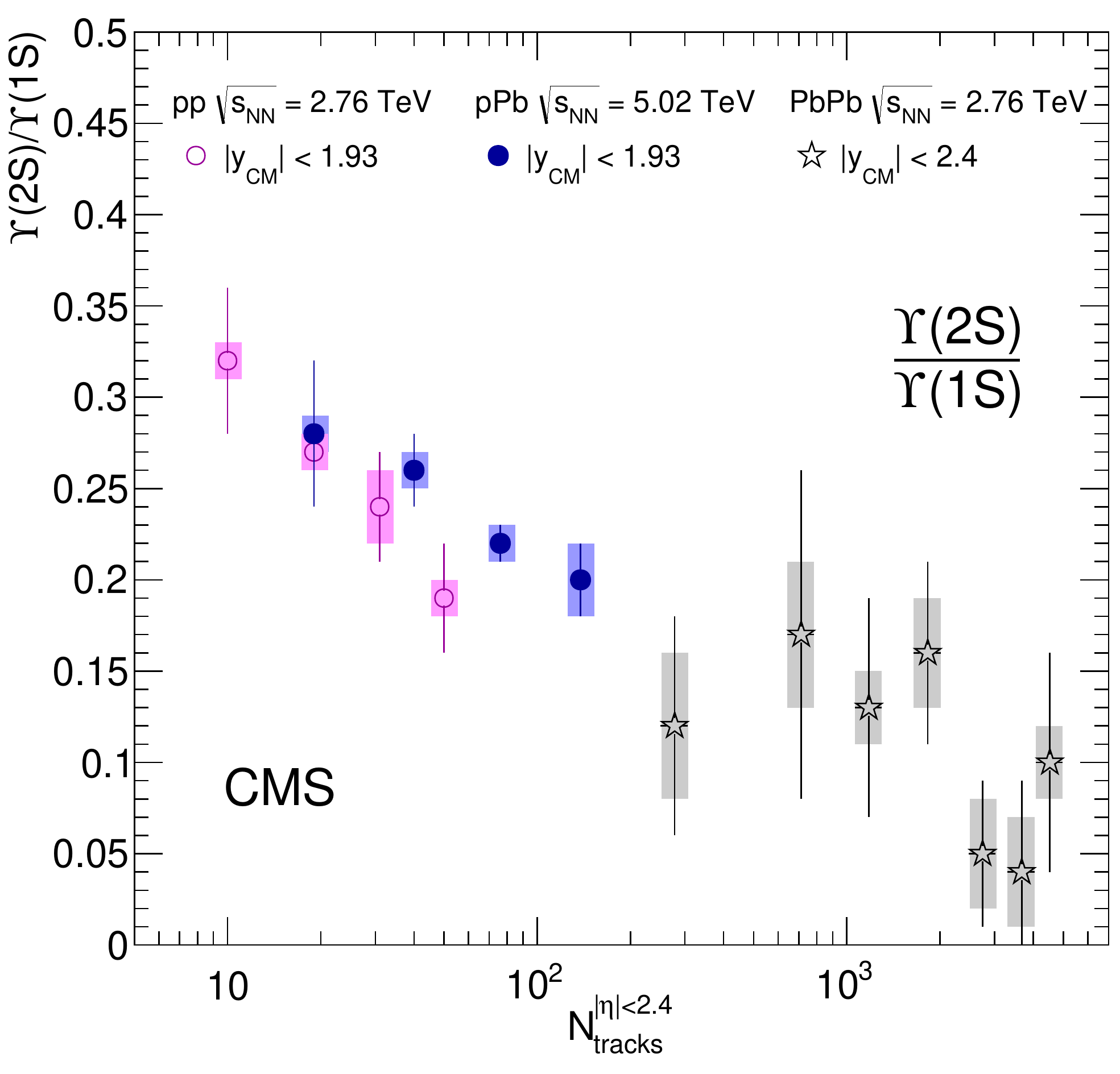}
    \caption{Single cross section ratios \singleRatioUpsBA for $\abs{y_{\mathrm{CM}}}<1.93$ versus (left) transverse energy measured at $4.0<\abs{\eta}<5.2$ and (right) charged-particle multiplicity measured in $\abs{\eta}<2.4$, for pp collisions at $\sqrts=2.76$\TeV (open circles) and \pPb collisions at $\sqrtsnn=5.02$\TeV (closed circles). Both figures also include the \singleRatioUpsBA ratios for $\abs{y_{\mathrm{CM}}}<2.4$ measured in \PbPb collisions at $\sqrtsnn=2.76$\TeV (open stars). The error bars in the figures indicate the statistical uncertainties, and the boxes represent the point-to-point systematic uncertainties. The global uncertainties of the results are 7\%, 8\%, and 8\% for the \pp, \pPb, and \PbPb, respectively. The results are available in tabulated form in Tables~\ref{tab:fig3} and \ref{tab:fig35aa}, with binning information provided in Tables~\ref{Tab:hfNtrkBin} and \ref{tab:fig35aa}.}
    \label{fig:ratioLogx}
  \end{center}
\end{figure}

The binned single ratios \singleRatioUpsBA and \singleRatioUpsCA are corrected for acceptance, and for trigger and reconstruction efficiencies. The bin-to-bin systematic uncertainties, represented by coloured boxes in Figs.~\ref{fig:ratio} and~\ref{fig:ratioLogx}, come from the fitting procedure and are in the ranges 3--8\% (\singleRatioUpsBA) and 4--30\% (\singleRatioUpsCA) for pp, and 3--8\% (\singleRatioUpsBA) and 7--17\% (\singleRatioUpsCA) for \pPb. The uncertainty common to all points in a given dataset, quoted in the captions, is estimated following the same procedure as for the activity-integrated results.

In Fig.~\ref{fig:ratio}, for both pp and \pPb, the results are shown as a function of forward transverse energy (\HFeta, left panel), and as a function of midrapidity track multiplicity (\Ntrks, right panel). In all bins, the abscissae are given by the bin-average value listed in Table~\ref{Tab:hfNtrkBin}. The ratios vary weakly as a function of \HFeta, while they exhibit a significant decrease with increasing \Ntrks.

The difference observed between the \PgU states when binning in \Ntrks can arise in two opposite ways. If, on the one hand, the \PgUa is systematically produced with more particles than the excited states, it would affect the underlying distribution of charged particles and create an artificial dependence when sliced in small multiplicity bins. This dependence should be sensitive to the underlying multiplicity distribution, and would result in a larger correlation if one reduces the size of the multiplicity bins. If, on the other hand, the $\Upsilon$ are interacting with the surrounding environment, the \PgUa is expected, as the most tightly bound state and the one of smallest size, to be less affected than $\PgUb$ and $\PgUc$, leading to a decrease of the $\PgUn/\PgUa$ ratios with increasing multiplicity. In either case, the ratios will continuously decrease from the pp to pPb to PbPb systems, as a function of event multiplicity.

The impact of additional underlying particles on the decreasing trend of the \singleRatioUpsBA and \singleRatioUpsCA versus \Ntrks in pp and pPb collisions is studied in more detail. The pp sample contains on average two extra charged tracks in the \PgUa events when compared to the \PgUb and \PgUc events, consistent with the \pPb sample, though the average number of charged particles rises from 13 (pp) to 50 (\pPb). The trend shown in the right panel of Fig.~\ref{fig:ratio} is found to weaken (or even reverse) if one artificially lowers the number of charged particles in the \PgUa sample by two or three tracks for every event. In contrast, the number of extra charged particles does not vary when lowering the \pt threshold down to 200$\MeVc$ in the \Ntrks computation, or when removing particles located in a cone of radius $\Delta R = \sqrt{\smash[b]{(\Delta\phi)^2+(\Delta\eta)^2}}= 0.3$ or 0.5 around the \PgU momentum direction. Extra charged particles are indeed expected in the \PgUa sample because of feed-down from higher-mass states, such as $\PgUb \to \PgUa  \pi^+ \pi^-$, but decay kinematics~\cite{Sjostrand:2006za}, with typically assumed feed-down fractions~\cite{Digal:2001ue}, do not lead to a significant rise of the number of charged particles with $\pt>400$\MeVc. While most feed-down contributions should come from the decays of P-wave states, such as $\chi_b \to \PgUa \gamma$, the probability for a photon to convert in the detector material and produce at least one electron with $\pt>400$\MeVc, that is further reconstructed and selected, is very low ($<$0.2\%). This makes the number of reconstructed electrons not sufficient to produce the measured trend. Therefore, it is concluded that feed-down contributions cannot solely account for the observed features in the measured ratios. It is noted also that if the three \PgU states are produced from the same initial partons, the mass difference between the \PgUa and the \PgUb ($>$500\MeV), or the \PgUa and the \PgUc ($>$800\MeV), could be found not only in the momentum of the \PgUa, but also in extra particles created together with the \PgUa.

For comparison, similarly corrected PbPb ratios, \singleRatioUpsBA, are computed from the double ratios presented in Ref.~\cite{Chatrchyan:2012lxa} versus percentiles of transverse energy deposited in the HF in the $2.9<\abs{\eta}<5.2$ range, which define the centrality of the PbPb event. The point-to-point systematic uncertainties are obtained as described in Ref.~\cite{Chatrchyan:2012lxa} and are in the range 13--85\% across all bins, while the 8\% global uncertainty is calculated as for the activity-integrated results described above. The statistical uncertainty ranges from 24\% to 139\%. Because there is a relatively strong correlation between the charged-particle multiplicity and the transverse energy in \PbPb collisions, the results reported here are not obtained by repeating the analysis as a function of \Ntrks, but by estimating, in the dimuon sample, the corresponding \Ntrks value for each of the HF energy-binned results~\cite{Chatrchyan:2012lxa}. The estimation is done using a low-multiplicity \PbPb sample reconstructed with the same reconstruction algorithm as the \pp and \pPb data, and the published \PbPb \pt charged-track distribution~\cite{Chatrchyan:2012ta} to account for the change in \pt shape between different \PbPb event activity categories.
Although the full HF acceptance is used for the centrality selection in \PbPb, the plotted transverse energy is scaled to the same pseudorapidity coverage as the \pp and \pPb datasets ($4.0<\abs{\eta}<5.2$) using the results in Ref.~\cite{Chatrchyan:2012mb}.

In Fig.~\ref{fig:ratioLogx}, the \singleRatioUpsBA ratios from the three collision systems are plotted versus \HFeta in the left panel, and versus \Ntrks in the right panel.
A logarithmic x-axis scale is chosen to allow displaying the three systems together. The relatively wide most peripheral (50--100\%) PbPb bin has little overlap with the highest-multiplicity pPb bin, preventing a direct comparison of the two systems at the same event activity. It should be noted that, within (large) uncertainties, the \PbPb centrality dependence is not pronounced~\cite{Chatrchyan:2012lxa} and that all \pp and \pPb ratios are far above the \PbPb activity-integrated ratio, shown in the right panel of Fig.~\ref{fig:dblRatio}.

\subsection{Self-normalized cross sections: \texorpdfstring{$\PgUn/\langle\PgUn\rangle$}{selfUpsilonnS}}
\begin{figure}[hbtp]
\centering
\includegraphics[width=0.32\textwidth]{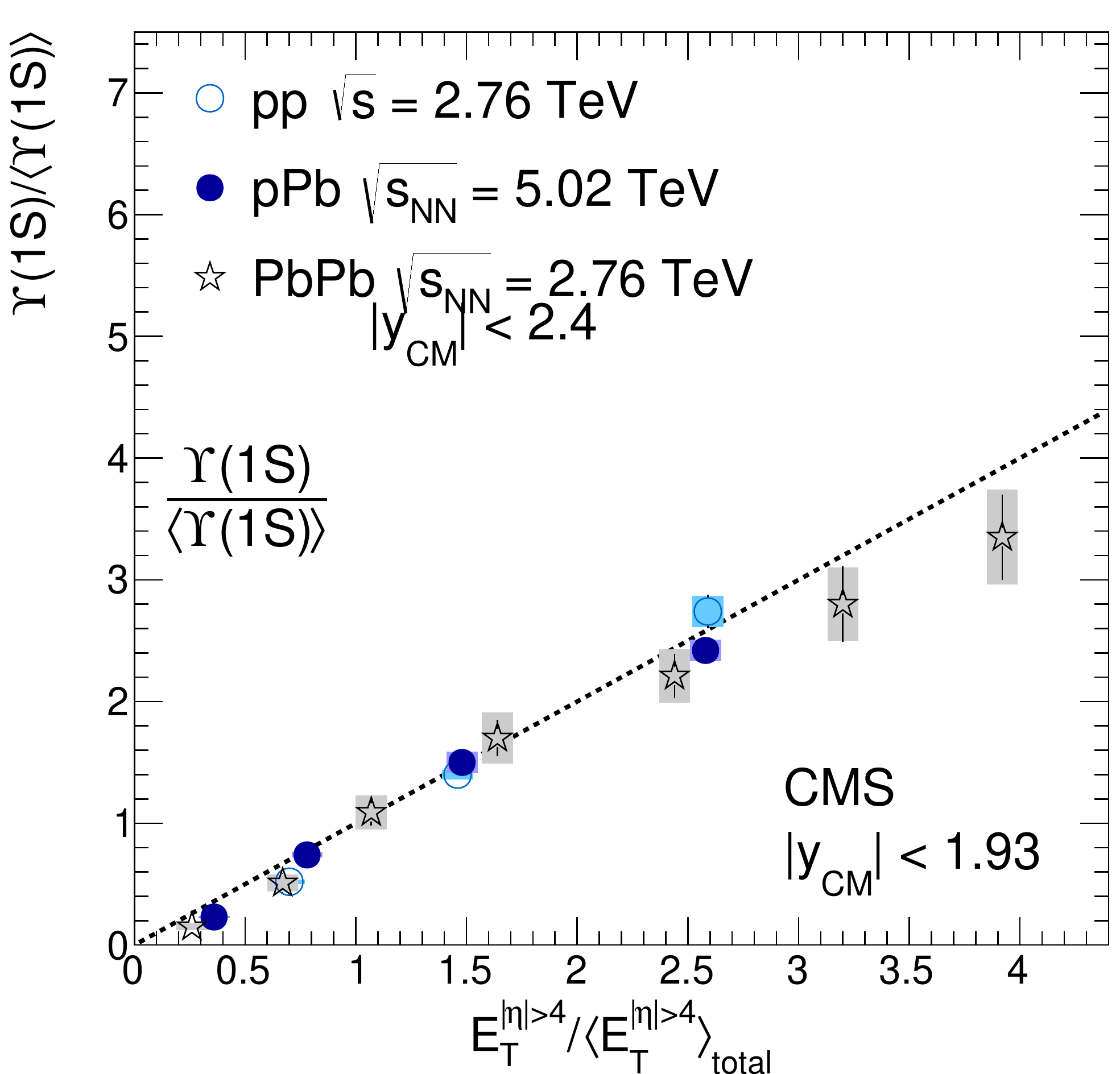}	\includegraphics[width=0.32\textwidth]{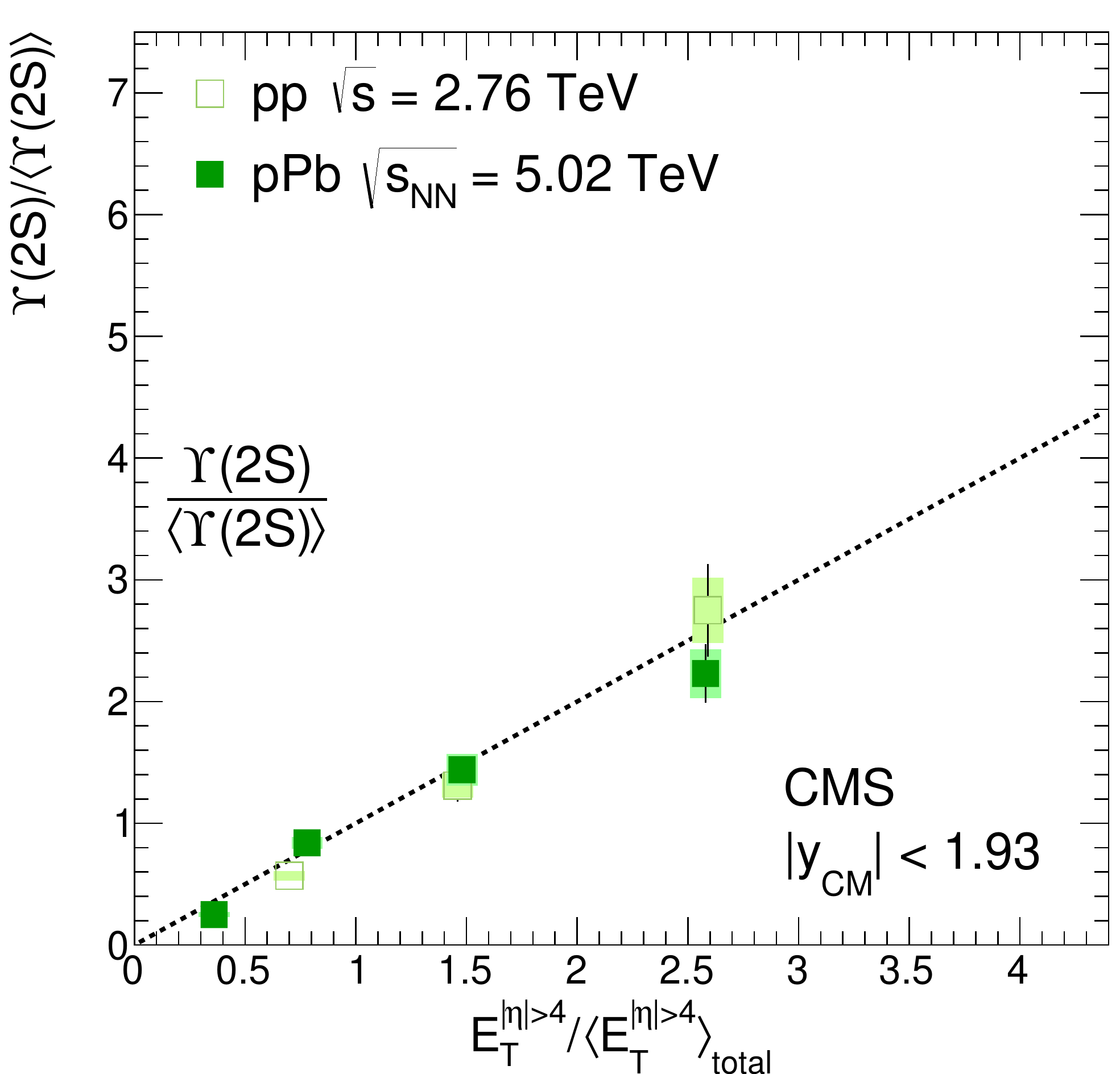}	\includegraphics[width=0.32\textwidth]{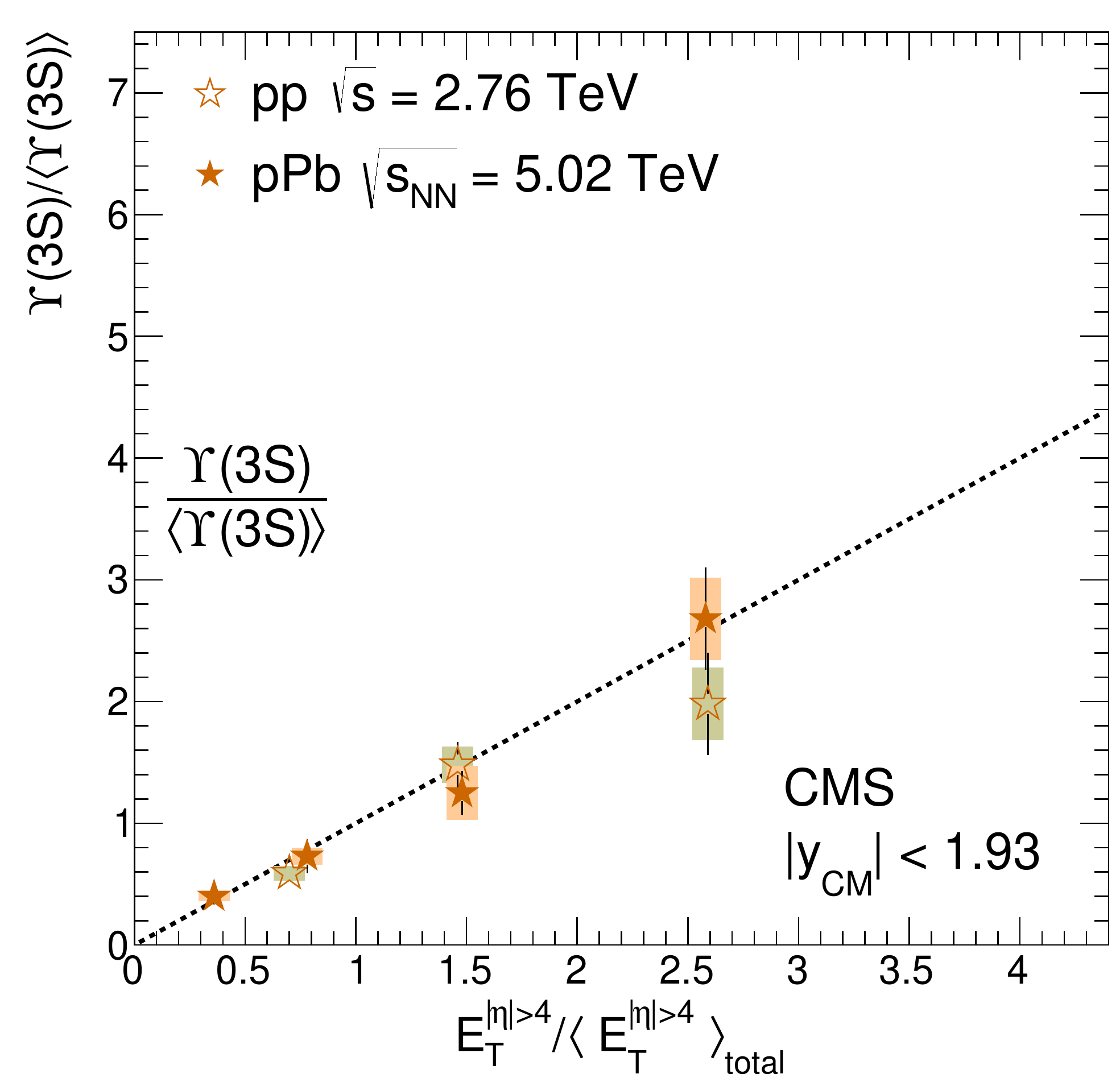}  	 \includegraphics[width=0.32\textwidth]{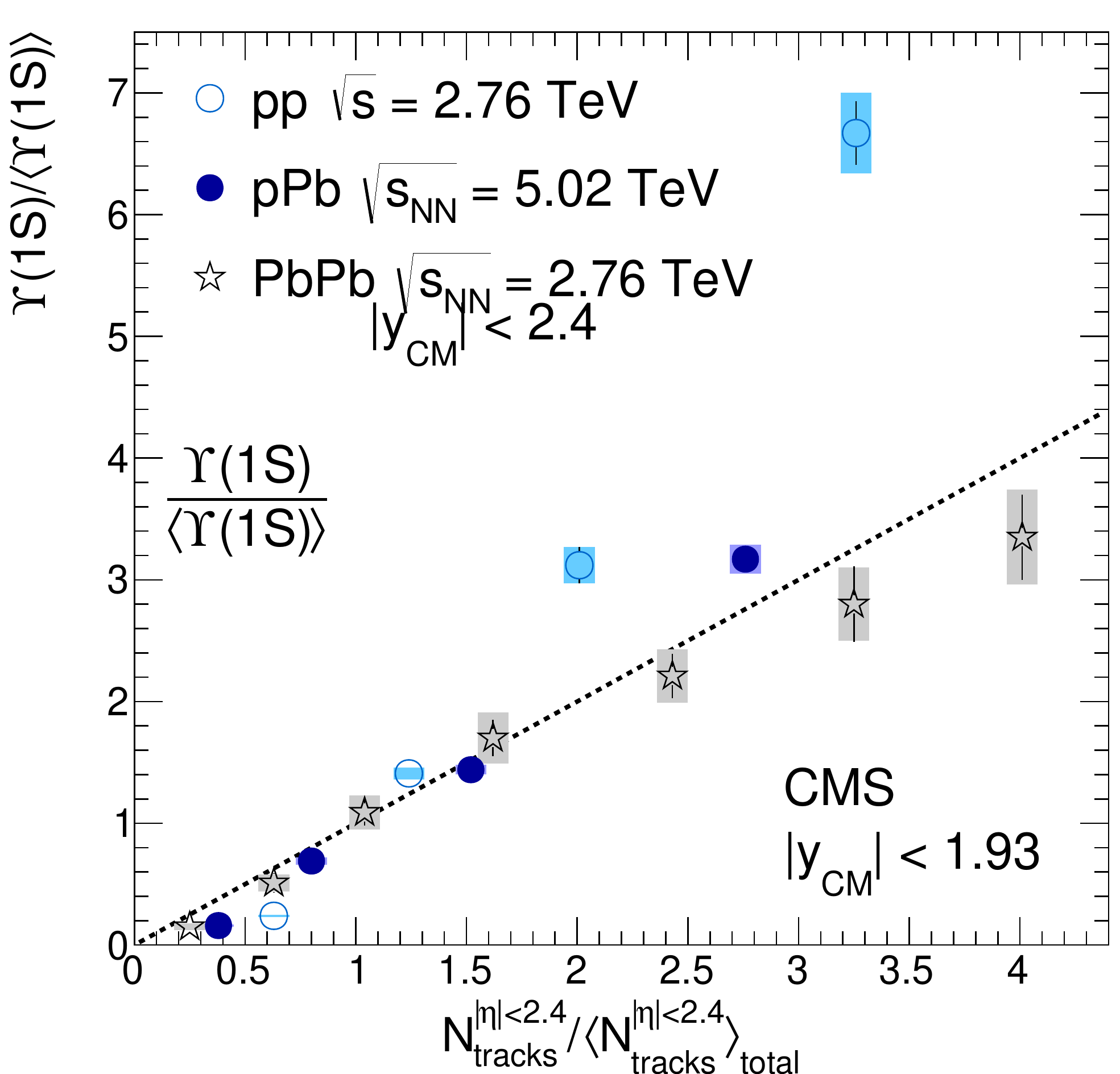}	 \includegraphics[width=0.32\textwidth]{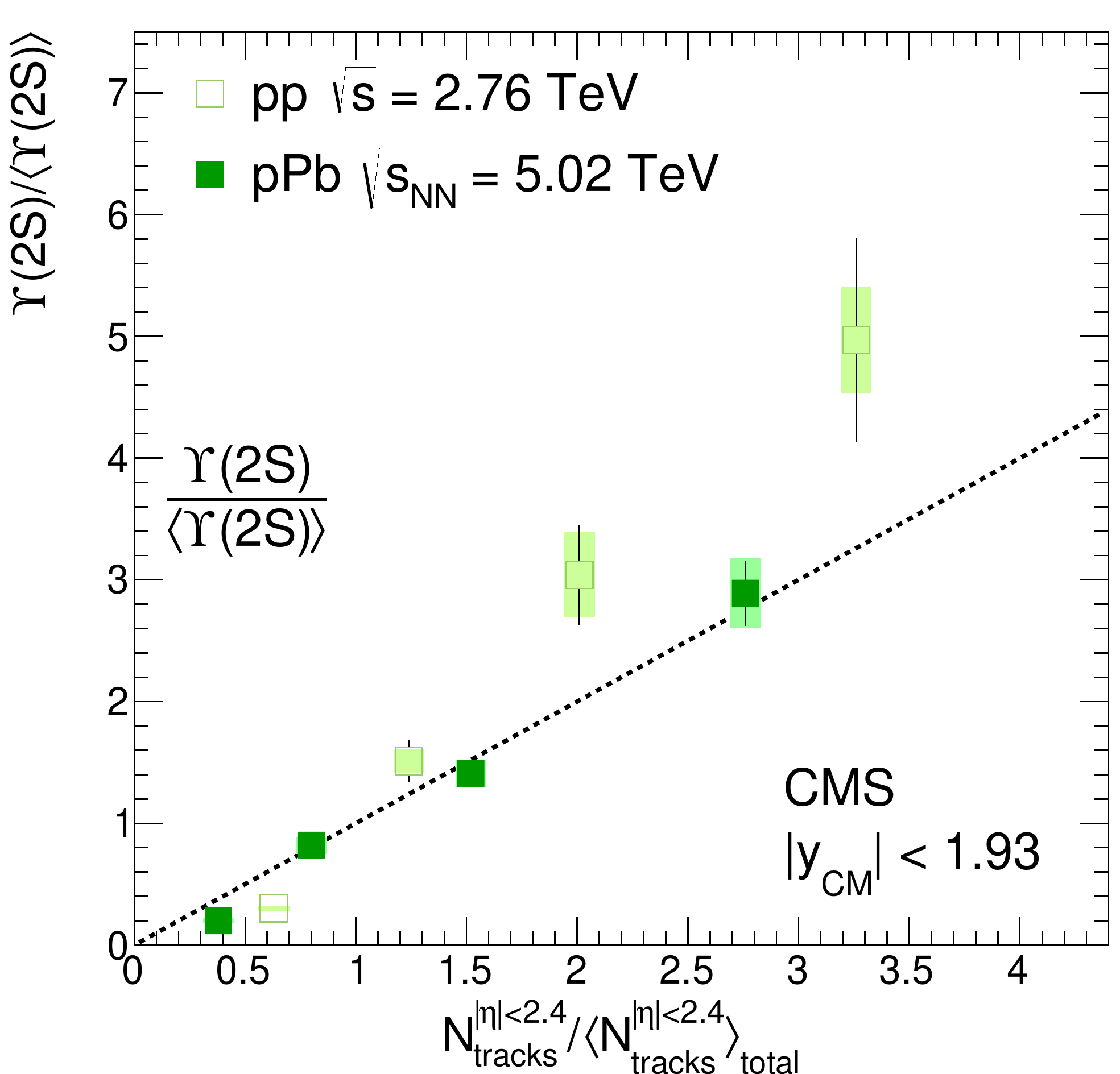} \includegraphics[width=0.32\textwidth]{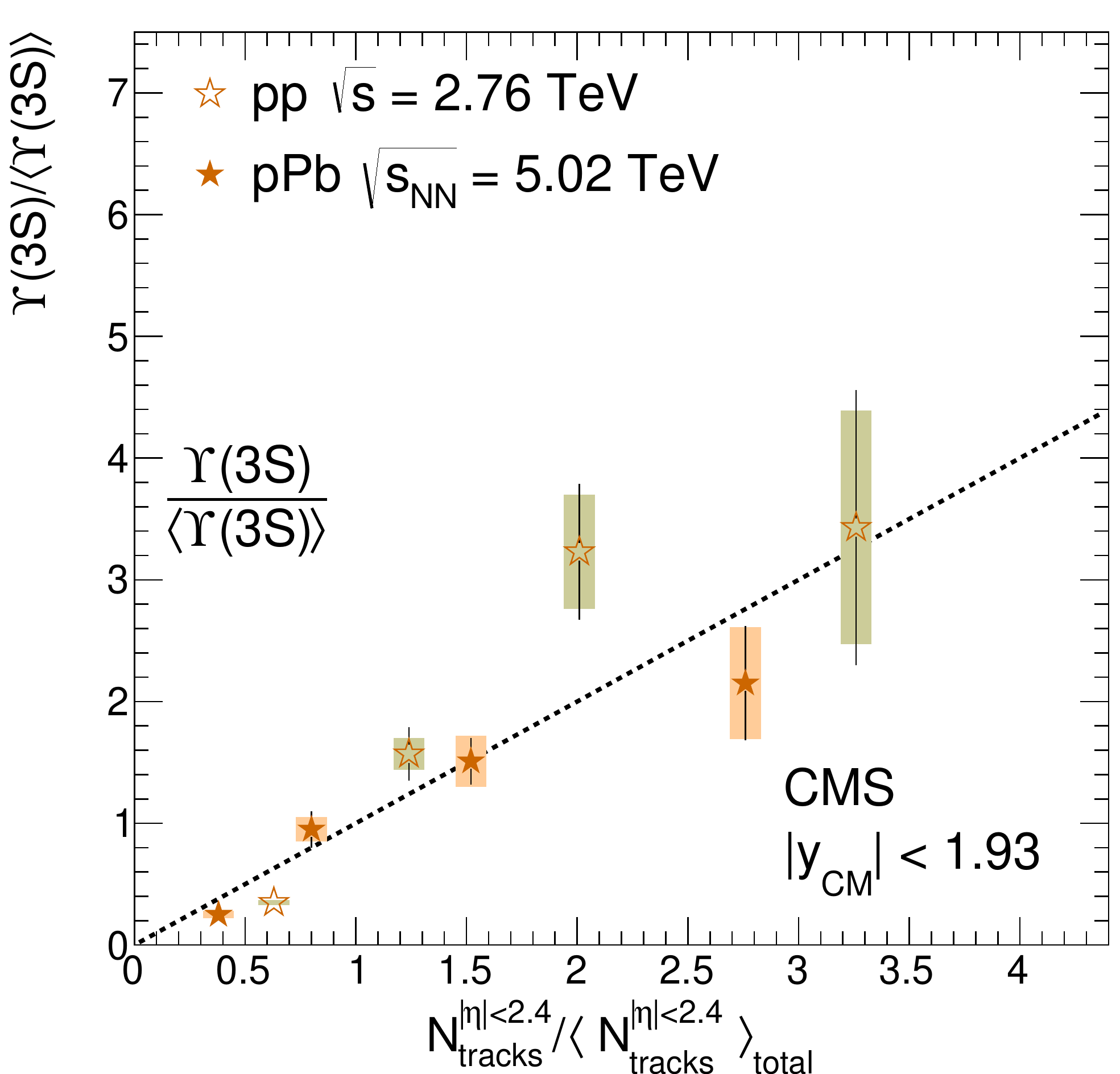}	
     \caption{The \PgUn cross section versus transverse energy measured at $4<\abs{\eta}<5.2$ (top row) and versus charged-track multiplicity measured in $\abs{\eta}<2.4$ (bottom row), measured in $\abs{y_{\mathrm{CM}}}<1.93$ in  pp collisions at $\sqrts=2.76$\TeV and \pPb collisions at $\sqrtsnn=5.02$\TeV. For \PgUa, the \PbPb data at $\sqrtsnn=2.76$\TeV (open stars) are overlaid. Cross sections and x-axis variables are normalized by their corresponding activity-integrated values. For all points, the abscissae are at the mean value in each bin. The dotted line is a linear function with a slope equal to unity. The error bars indicate the statistical uncertainties, and the boxes represent the point-to-point systematic uncertainties. The results are available in tabulated form in Table~\ref{tab:fig5}.}
    \label{fig:yields}
\end{figure}

All the ratios presented so far address the relative differences between the excited states and the ground state. In addition, the individual \PgUn yields, self-normalized to their activity-integrated values, are computed. The results are shown in Fig.~\ref{fig:yields} in bins of \HFetaRelative (top) and \NtrksRelative (bottom), for pp and \pPb collisions, where the denominator is averaged over all events. These ratios are constructed from the yields extracted from the same fit as the single ratios and are corrected for the residual activity-dependent efficiency that does not cancel in the ratio. The systematic uncertainties are determined following the same procedure as for the other results reported in this paper.
The bin-to-bin systematic uncertainties, represented by the coloured boxes in Fig.~\ref{fig:yields}, come from the fitting procedure and are in the ranges 3--7\% (\PgUa), 5--14\% (\PgUb) and 6--20\% (\PgUc), depending on the bin. Figure~\ref{fig:yields} (left) also shows the corresponding ratios for the \PgUa state in \PbPb collisions, which are derived from Ref.~\cite{Chatrchyan:2012lxa} by dividing the nuclear modification factors (\raa) binned in centrality by the centrality-integrated \raa value. The \PgUb results from Ref.~\cite{Chatrchyan:2012lxa} are not included here because of their low precision.

All the self-normalized cross section ratios increase with increasing forward transverse energy and midrapidity particle multiplicity in the event. In the cases where Pb ions are involved, the increase observed in both variables can arise from the increase in the number of nucleon-nucleon collisions. The pp results are reminiscent of a similar \JPsi measurement made in pp collisions at 7\TeV~\cite{Abelev:2012rz}. A possible interpretation of the positive correlation between the \PgU production yield and the underlying activity of the pp event is the occurrence of multiple parton-parton interactions in a single pp collision~\cite{Abramowicz:2013iva}.

To compare the trends between collision systems, linear fits (not shown) are performed separately for the pp, \pPb, and \PbPb results. In the case of the forward transverse energy binning, the self-normalized ratios in all three collision systems are found to have a slope consistent with unity. Hence, no significant difference between \pp, \pPb, and \PbPb results or between individual states is observed when correlating \PgU production yields with forward event activity. The similarity of the three systems has to be tempered by the fact that very different mean values are used for normalizing the forward transverse energy, 3.5, 14.7, and 765 GeV, respectively, as well as by the absence of sensitivity of the $\PgUn/\langle\PgUn\rangle$ observable to a modification that is independent of event activity. In contrast, the case of \Ntrks binning shows differences between the three states, an observation which is related to the single-ratio variations observed in Fig.~\ref{fig:ratio} (right). The \PgUa, in particular, exhibits the fastest rise in pp collisions.

\section{Summary}
The relative production of the three \PgU states has been investigated in \pPb and pp collisions collected in 2013 by the CMS experiment, in the $\abs{y_{\mathrm{CM}}}<1.93$ centre-of-mass rapidity range. The self-normalized cross section ratios, $\PgUa/\langle\PgUa\rangle$, $\PgUb/\langle\PgUb\rangle$, $\PgUc/\langle\PgUc\rangle$, increase with event activity. The excited-to-ground-states cross section ratios, $\Upsilon\text{(nS)}/\Upsilon\text{(1S)}$, are found to decrease with increasing charged-particle multiplicity as measured in the $\abs{\eta}<2.4$ pseudorapidity interval that contains the region in which the \PgU are measured. This unexpected dependence suggests novel phenomena in quarkonium production that could arise from a larger number of charged particles being systematically produced with the ground state, or from a stronger impact of the growing number of nearby particles on the more weakly bound states. This dependence is less pronounced when the event activity is inferred from transverse energy deposited in the forward $4.0 < \abs{\eta} < 5.2$ region. When integrated over event activity, the double ratios $[\Upsilon\text{(nS)}/\Upsilon\text{(1S)}]_{\pPb} / [\Upsilon\text{(nS)}/\Upsilon\text{(1S)}]_{\pp}$ are found to be equal to $0.83 \pm 0.05\stat\pm0.05\syst$ and $0.71 \pm 0.08\stat\pm0.09\syst$ for $\Upsilon\text{(2S)}$ and $\Upsilon\text{(3S)}$, respectively, which are larger than the corresponding double ratios measured for \PbPb collisions. This suggests the presence of final-state suppression effects in the \pPb collisions compared to pp collisions which affect more strongly the excited states (\PgUb and \PgUc) compared to the ground state (\PgUa). A global understanding of the effects at play in pp, pPb, and PbPb calls for more activity-related studies of the $\Upsilon$ yields in pp collisions, as well as for additional PbPb data allowing a more detailed investigation of the most peripheral events.

We congratulate our colleagues in the CERN accelerator departments for the excellent performance of the LHC machine. We thank the technical and administrative staff at CERN and other CMS institutes, and acknowledge support from: BMWF and FWF (Austria); FNRS and FWO (Belgium); CNPq, CAPES, FAPERJ, and FAPESP (Brazil); MES (Bulgaria); CERN; CAS, MoST, and NSFC (China); COLCIENCIAS (Colombia); MSES (Croatia); RPF (Cyprus); MEYS (Czech Republic); MoER, SF0690030s09 and ERDF (Estonia); Academy of Finland, MEC, and HIP (Finland); CEA and CNRS/IN2P3 (France); BMBF, DFG, and HGF (Germany); GSRT (Greece); OTKA and NKTH (Hungary); DAE and DST (India); IPM (Iran); SFI (Ireland); INFN (Italy); NRF and WCU (Korea); LAS (Lithuania); CINVESTAV, CONACYT, SEP, and UASLP-FAI (Mexico); MSI (New Zealand); PAEC (Pakistan); MSHE and NSC (Poland); FCT (Portugal); JINR (Armenia, Belarus, Georgia, Ukraine, Uzbekistan); MON, RosAtom, RAS and RFBR (Russia); MSTD (Serbia); SEIDI and CPAN (Spain); Swiss Funding Agencies (Switzerland); NSC (Taipei); TUBITAK and TAEK (Turkey); NASU (Ukraine); STFC (United Kingdom); DOE and NSF (USA). Individuals have received support from the Marie-Curie programme and the European Research Council and EPLANET (European Union); the Leventis Foundation; the A. P. Sloan Foundation; the Alexander von Humboldt Foundation; the Belgian Federal Science Policy Office; the Fonds pour la Formation \`a la Recherche dans l'Industrie et dans l'Agriculture (FRIA-Belgium); the Agentschap voor Innovatie door Wetenschap en Technologie (IWT-Belgium); the Ministry of Education, Youth and Sports (MEYS) of Czech Republic; the Council of Science and Industrial Research, India; the Compagnia di San Paolo (Torino); the HOMING PLUS programme of Foundation for Polish Science, cofinanced by EU, Regional Development Fund; and the Thalis and Aristeia programmes cofinanced by EU-ESF and the Greek NSRF.

\bibliography{auto_generated}   

\appendix
\section{Results in tabulated format\label{app:datatables}}

\begin{table}[h]
  \begin{center}
    \topcaption{The $\sqrt{s}$ dependence of the excited-to-ground-state cross section ratios, $\frac{\PgUn}{\PgUa}$, in \pp and \ppbar collisions. The total quoted uncertainties represent the quadratic sum of the statistical, systematic, and global uncertainties. Listed also are the $\Upsilon$ rapidity and transverse momentum ranges for which each measurement is reported.}
    \label{tab:sqrts}
    \begin{tabular}{lllcc}
      \hline
       Data & \pt [\GeVcns{}] & Rapidity & $\frac{\Upsilon(2S)}{\Upsilon(1S)}$ total &$\frac{\Upsilon(3S)}{\Upsilon(1S)}$ total\\
      \hline
				     CMS \pp $\sqrts=2.76$\TeV                                             & 0--40  &  $\abs{y}<1.93$  &   $0.26 \pm 0.02 $ & $0.11 \pm 0.02 $\\
      			             CMS \pp $\sqrts=7$\TeV~\cite{Chatrchyan:2013yna}      & 0--38  &  $\abs{y}<2.4$    &   $0.26 \pm 0.03 $ & $0.13 \pm0.02$\\
				     CDF \ppbar $\sqrts=1.9$\TeV~\cite{Abe:1995an}            & 1--10  &  $\abs{y}<0.4$    &   $0.28 \pm 0.05 $ & $0.16 \pm 0.03 $\\
      \hline	
     \end{tabular}

  \end{center}
\end{table}
\begin{table}[h]
  \begin{center}
    \topcaption{The excited-to-ground-state cross section ratios, $\frac{\PgUn}{\PgUa}$, for Upsilons with $\pt<40$\GeVc, in \pp, \pPb, and \PbPb collisions at nucleon-nucleon center of mass collision energy of 2.76, 5.02, and 2.76\TeV, respectively. Listed uncertainties are statistical first, systematic second, and global third.}
    \label{tab:fig2left}
    \begin{tabular}{llcc}
      \hline
       Data & Rapidity & $\frac{\Upsilon(2S)}{\Upsilon(1S)}$ &$\frac{\Upsilon(3S)}{\Upsilon(1S)}$\\
      \hline
				     \pp    $\sqrts=2.76$\TeV         &  $\abs{y_{\mathrm{CM}}}<1.93$   &   $0.26 \pm 0.01 \pm 0.01 \pm 0.02 $ & $0.11 \pm 0.01 \pm 0.01 \pm 0.01$\\
      			             \pPb  $\sqrtsnn=5.02$\TeV     &  $\abs{y_{\mathrm{CM}}}<1.93$   &   $0.22 \pm 0.01 \pm 0.01 \pm 0.02 $ & $0.08 \pm0.01 \pm 0.01 \pm 0.01$\\
				    \PbPb $\sqrtsnn=2.76$\TeV    &  $\abs{y_{\mathrm{CM}}}<2.4$     &   $0.09 \pm 0.02 \pm 0.02 \pm 0.01 $ & $<$0.04 (at 95\% confidence level)\\
      \hline	
     \end{tabular}

  \end{center}
\end{table}

\begin{table}[htbH]
\begin{center}
\topcaption{Event activity bins in \Ntrks (left) and \HFeta (right), comprising the bin edges, the mean within the bin and the corresponding mean of the other variable calculated in the dimuon sample, and the fraction of recorded minimum bias triggered events falling within the bin. The bin upper boundaries are chosen for each variable so that they are half or round multiples of the uncorrected mean value in the minimum bias events, $ {\langle N_{\text{tracks, raw}}^{\abs{\eta}<2.4}\rangle}= 10$ and 41, ${\langle E_{\text{T, raw}}^{\abs{\eta}>4} \rangle}= 3.5$ and 14.7\GeV for pp and \pPb, respectively. The quoted $\langle \Ntrks \rangle$ values are efficiency corrected.}
\label{Tab:hfNtrkBin}
\begin{tabular}{cccccc|cccc}

\hline
& Bin &\multicolumn{4}{c|}{\Ntrks} & \multicolumn{4}{c}{\HFeta}\\
&&&&&&&&&\\
\hline
&& [\Ntrks]  & $\langle \Ntrks \rangle$ & $\langle\HFeta \rangle$& Frac  & [\HFeta] & $\langle \HFeta \rangle$& $\langle\Ntrks \rangle$ & Frac\\
&& (raw)    &                          & [\GeVns{}]                            & (\%)   & [\GeVns{}]    &                                        &                                      & (\%)\\
&&&&&&&&\\
\hline

\multirow{4}{2em}{pp}
 & 1 & 0--10         &  $9.8\pm 0.4$    &  3.3   &  64   & 0--3.5          & 2.5    &  $9.6\pm 0.4$ & 59 \\
 & 2 & 11--20       &  $19.4\pm 0.8$  &  4.7   &  25   & 3.5--7.0       &  5.2   &  $17.2\pm 0.7$ & 32 \\
 & 3 & 21--30       &  $30.7\pm 1.2$  &  5.9   &  8     & $\geq$7.0   & 9.2    &   $25.8\pm 1.0$ & 9  \\
 & 4 & $\geq$31  &  $49.9\pm 1.9$  &  7.1   &  3     &   &    &    &   \\

\hline
\multirow{4}{2em}{\pPb}
                     & 1 & 0--21          & $19.1\pm 0.7$   &    7.3   & 35   &           0--7.4      &   5.3   &  $19.2\pm0.7$   & 30  \\
                     & 2 & 22--41        & $40.0\pm 1.6$   &  13.0  & 24   &      7.4--14.7      &  11.5  & $40.2\pm 1.6$    & 27  \\
                     & 3 & 42--82       & $75.9\pm 3.0$   &  21.6  & 30   &     14.7--29.4     &  21.8  & $72.8\pm 2.8$    & 33  \\
                     & 4 & $\geq$83  & $137.9\pm 5.4$  & 34.4  & 11    &    $\geq$29.4   &  38.0  & $118.0\pm 4.6$  & 10 \\

\hline

\end{tabular}
\end{center}
\end{table}

\begin{table}[h]
  \begin{center}
    \topcaption{Excited-to-ground state cross section ratios, in event activity bins. Listed uncertainties are statistical
      first, systematic second, and global scale third.}
    \label{tab:fig3}
    \begin{tabular}{cccc}
          \hline
         & Bin & $\frac{\PgUb}{\PgUa}$ &$\frac{\PgUc}{\PgUa}$ \\
      \hline
      \multicolumn{4}{c}{\HFeta} \\
      \hline
      \multirow{3}{2em}{pp}
				     & 1 &  $0.27  \pm 0.03 \pm 0.01 \pm 0.02$  &    $0.12 \pm 0.02 \pm 0.01 \pm 0.01$\\
      			              & 2 &  $0.23 \pm 0.02 \pm 0.01 \pm 0.02$  &    $0.12 \pm 0.01 \pm 0.01 \pm 0.01$\\
                                       & 3 &  $0.25 \pm 0.03 \pm 0.01 \pm 0.02$  &    $0.08 \pm 0.02 \pm 0.01 \pm 0.01$\\
      \hline	
      \multirow{4}{2em}{pPb}
      				      & 1 &  $0.25 \pm 0.04 \pm 0.01 \pm 0.02$  &    $0.13 \pm 0.03 \pm 0.01 \pm 0.01$\\
      			              & 2 &  $0.25 \pm 0.02 \pm 0.01 \pm 0.02$  &    $0.07 \pm 0.01 \pm 0.01 \pm 0.01$\\
                                       & 3 &  $0.22 \pm 0.01 \pm 0.01 \pm 0.02$  &    $0.06 \pm 0.01 \pm 0.01 \pm 0.01$\\
                                       & 4 &  $0.21 \pm 0.02 \pm 0.01 \pm 0.02$  &    $0.09 \pm 0.01 \pm 0.01 \pm 0.01$\\
      \hline	
         \multicolumn{4}{c}{\Ntrks} \\
      \hline
      \multirow{4}{2em}{pp}
				      & 1 &  $0.32 \pm 0.04 \pm 0.01 \pm 0.02$  &    $0.16 \pm 0.02 \pm 0.01 \pm 0.01$\\
      			              & 2 &  $0.27 \pm 0.02 \pm 0.01 \pm 0.02$  &    $0.12 \pm 0.01 \pm 0.01 \pm 0.01$\\
                                       & 3 &  $0.24 \pm 0.03 \pm 0.02 \pm 0.02$  &    $0.11 \pm 0.02 \pm 0.01 \pm 0.01$\\
                                       & 4 &  $0.19 \pm 0.03 \pm 0.01 \pm 0.01$  &    $0.06 \pm 0.02 \pm 0.02 \pm 0.00$\\
      \hline	
      \multirow{4}{2em}{pPb}
      				      & 1 &  $0.28 \pm 0.04 \pm 0.01 \pm 0.02$  &    $0.12 \pm 0.03 \pm 0.01 \pm 0.01$\\
      			              & 2 &  $0.26 \pm 0.02 \pm 0.01 \pm 0.02$  &    $0.10 \pm 0.02 \pm 0.01 \pm 0.01$\\
                                       & 3 &  $0.22 \pm 0.01 \pm 0.01 \pm 0.02$  &    $0.08 \pm 0.01 \pm 0.01 \pm 0.01$\\
                                       & 4 &  $0.20 \pm 0.02 \pm 0.02 \pm 0.02$  &    $0.05 \pm 0.01 \pm 0.01 \pm 0.00$\\
      \hline	

     \end{tabular}
  \end{center}
\end{table}

\begin{table}[h]
  \begin{center}
    \topcaption{Self-normalized cross section ratios, in event activity bins. In the first column for each bin, the numerator is averaged over the bin and the denominator is averaged over all events. Listed uncertainties are statistical first and systematic second.}
    \label{tab:fig5}
    \begin{tabular}{cccccc}
          \hline
        & Bin &  & $\frac{\PgUa}{\langle\PgUa\rangle}$ &$\frac{\PgUb}{\langle\PgUb\rangle}$ & $\frac{\PgUc}{\langle\PgUc\rangle}$\\
      \hline
      &&$\frac{\langle\HFeta\rangle}{{\langle\HFeta\rangle}_{\text{total}}}$&\multicolumn{3}{c}{\HFeta} \\
      \hline
      \multirow{3}{2em}{pp}
				      &  1   &  0.70   & $0.52  \pm 0.02 \pm 0.02 $  &   $0.57 \pm 0.07 \pm 0.04 $   &   $0.59 \pm 0.08 \pm 0.06 $\\
      			              &  2   &  1.46   & $1.40 \pm 0.05 \pm 0.04 $   &    $1.31 \pm 0.13 \pm 0.10 $  &   $1.48 \pm 0.19 \pm 0.15 $\\
                                       &  3   &   2.59  & $2.74 \pm 0.14 \pm 0.13 $   &    $2.75 \pm 0.38 \pm 0.27 $  &   $1.98 \pm 0.42 \pm 0.30 $\\
      \hline	
      \multirow{4}{2em}{pPb}
      				      & 1 &  0.36   & $0.23 \pm 0.01 \pm 0.01$  &    $0.25 \pm 0.04 \pm 0.02$  &   $0.40 \pm 0.07 \pm 0.04 $\\
      			              & 2 &  0.78   & $0.74 \pm 0.03 \pm 0.02$  &    $0.84 \pm 0.07 \pm 0.05$  &   $0.73 \pm 0.14 \pm 0.07 $\\
                                       & 3 &  1.48   & $1.50 \pm 0.04 \pm 0.09$  &    $1.44 \pm 0.10 \pm 0.13$  &   $1.25 \pm 0.18 \pm 0.22 $\\
                                       & 4 &  2.58   & $2.42 \pm 0.08 \pm 0.09$  &    $2.23 \pm 0.24 \pm 0.20$  &   $2.68 \pm 0.42 \pm 0.34 $\\
      \hline	
        &&$\frac{\langle\Ntrks\rangle}{{\langle\Ntrks\rangle}_{\text{total}}}$& \multicolumn{3}{c}{\Ntrks} \\
      \hline
      \multirow{4}{2em}{pp}
 				      & 1 &  0.63  & $0.24  \pm 0.01 \pm 0.01 $  &   $0.30 \pm 0.05 \pm 0.02 $  &   $0.35 \pm 0.05 \pm 0.02 $\\
      			              & 2 &  1.24  & $1.41 \pm 0.06 \pm 0.05 $   &    $1.51 \pm 0.17 \pm 0.10 $ &   $1.57 \pm 0.22 \pm 0.13 $\\
                                       & 3 &  2.01  & $3.12 \pm 0.15 \pm 0.15 $   &    $3.04 \pm 0.41 \pm 0.35 $ &   $3.23 \pm 0.56 \pm 0.47 $\\
                                       & 4 &  3.26  & $6.67 \pm 0.26 \pm 0.33$    &    $4.97 \pm 0.84 \pm 0.44 $ &   $3.43 \pm 1.13 \pm 0.96 $\\
      \hline	
      \multirow{4}{2em}{pPb}
      				      & 1 &  0.38  & $0.16 \pm 0.01 \pm 0.01$  &    $0.20 \pm 0.03 \pm 0.02$ &    $0.25 \pm 0.05 \pm 0.03$\\
      			              & 2 &  0.80  & $0.69 \pm 0.03 \pm 0.03$  &    $0.82 \pm 0.09 \pm 0.07$ &    $0.95 \pm 0.15 \pm 0.10$\\
                                       & 3 &  1.52  & $1.44 \pm 0.04 \pm 0.04$  &    $1.41 \pm 0.11 \pm 0.11$  &    $1.51 \pm 0.19 \pm 0.21$\\
                                       & 4 &  2.76  & $3.17 \pm 0.09 \pm 0.12$  &    $2.89 \pm 0.27 \pm 0.29$ &    $2.15 \pm 0.47 \pm 0.46$\\
      \hline	

     \end{tabular}
  \end{center}
\end{table}

\begin{table}[h]
  \begin{center}
    \topcaption{Single cross section ratios, \singleRatioUpsBA and $\PgUa$/$\langle\PgUa\rangle$, measured in bins of centrality (Cent.) in \PbPb collisions at $\sqrtsnn=2.76$\TeV, derived from Ref.~\cite{Chatrchyan:2012lxa}. The quoted $\langle \Ntrks \rangle$ values are efficiency corrected. In the second section, the denominator in the fractions is averaged over all events. Listed uncertainties are statistical first, systematic second, and global scale third.}
    \label{tab:fig35aa}
    \begin{tabular}{lllc}
      \hline
       Cent. & $\langle\Ntrks\rangle$  & $\langle\HFeta\rangle$ &  $\frac{\PgUb}{\PgUa}$\\
                &                 & [\GeVns{}]                        &  \\
      \hline
        100--50\%  &  $278\pm28$     &  77      & $0.12 \pm 0.06 \pm 0.04 \pm 0.01$ \\
        50--40\%    &  $712\pm71$     &  192    & $0.17 \pm 0.09 \pm 0.04 \pm 0.01$ \\
        40--30\%    &  $1178\pm118$ &  302   & $0.13 \pm 0.06 \pm 0.02 \pm 0.01$ \\
        30--20\%    &  $1825\pm183$ &  459   & $0.16 \pm 0.05 \pm 0.03 \pm 0.01$ \\
        20--10\%    &  $2744\pm274$ &  681   & $0.05 \pm 0.04 \pm 0.03 \pm 0.01$ \\
        10--5\%      &  $3672\pm367$ &  892   & $0.04 \pm 0.05 \pm 0.03 \pm 0.01$ \\
        5--0\%        &  $4526\pm453$ &  1093 & $0.10 \pm 0.06 \pm 0.02 \pm 0.01$ \\
        \hline	
        Cent. & $\frac{\langle\Ntrks\rangle}{{\langle\Ntrks\rangle}_{\text{total}}}$  &$\frac{\langle\HFeta\rangle}{{\langle\HFeta\rangle}_{\text{total}}}$ &  $\frac{\PgUa}{\langle\PgUa\rangle}$\\
        \hline	
       100--50\%   &  0.25  &  0.26   & $0.15 \pm 0.02 \pm 0.03$ \\
        50--40\%    &  0.63  &  0.67   & $0.51 \pm 0.08 \pm 0.07$ \\
        40--30\%    &  1.04  &  1.07   & $1.09 \pm 0.11 \pm 0.14$ \\
        30--20\%    &  1.62  &  1.64   & $1.70 \pm 0.15 \pm 0.21$ \\
        20--10\%    &  2.43  &  2.44   & $2.21 \pm 0.18 \pm 0.22$ \\
        10--5\%      &  3.25  &  3.20   & $2.80 \pm 0.31 \pm 0.30$ \\
        5--0\%        &  4.01  &  3.92   & $3.35 \pm 0.35 \pm 0.39$ \\

      \hline	
     \end{tabular}

  \end{center}
\end{table}
\cleardoublepage \section{The CMS Collaboration \label{app:collab}}\begin{sloppypar}\hyphenpenalty=5000\widowpenalty=500\clubpenalty=5000\textbf{Yerevan Physics Institute,  Yerevan,  Armenia}\\*[0pt]
S.~Chatrchyan, V.~Khachatryan, A.M.~Sirunyan, A.~Tumasyan
\vskip\cmsinstskip
\textbf{Institut f\"{u}r Hochenergiephysik der OeAW,  Wien,  Austria}\\*[0pt]
W.~Adam, T.~Bergauer, M.~Dragicevic, J.~Er\"{o}, C.~Fabjan\cmsAuthorMark{1}, M.~Friedl, R.~Fr\"{u}hwirth\cmsAuthorMark{1}, V.M.~Ghete, C.~Hartl, N.~H\"{o}rmann, J.~Hrubec, M.~Jeitler\cmsAuthorMark{1}, W.~Kiesenhofer, V.~Kn\"{u}nz, M.~Krammer\cmsAuthorMark{1}, I.~Kr\"{a}tschmer, D.~Liko, I.~Mikulec, D.~Rabady\cmsAuthorMark{2}, B.~Rahbaran, H.~Rohringer, R.~Sch\"{o}fbeck, J.~Strauss, A.~Taurok, W.~Treberer-Treberspurg, W.~Waltenberger, C.-E.~Wulz\cmsAuthorMark{1}
% ...
\vskip\cmsinstskip
% ...
1:~~Also at Vienna University of Technology, Vienna, Austria\\
2:~~Also at CERN, European Organization for Nuclear Research, Geneva, Switzerland\\
% ...

\end{sloppypar}
\end{document}